# Anomalous transport due to Weyl fermions in the chiral antiferromagnets $Mn_3X$, $X$ = Sn, Ge

Taishi Chen[1,2*], Takahiro Tomita[2,3*], Susumu Minami[1,4,5], Mingxuan Fu[1,2*], Takashi Koretsune[6], Motoharu Kitatani[4], Ikhlas Muhammad[2], Daisuke Nishio-Hamane[2], Rieko Ishii[2], Fumiyuki Ishii[4,5], Ryotaro Arita[3,4,7] & Satoru Nakatsuji[1,2,3,8,9]

[1]*Department of Physics, University of Tokyo, Tokyo, Japan*

[2]*Institute for Solid State Physics, University of Tokyo, Kashiwa, Chiba, Japan*

[3]*CREST, Japan Science and Technology Agency (JST), Honcho Kawaguchi, Japan*

[4]*RIKEN Center for Emergent Matter Science (CEMS), Wako, Saitama, Japan*

[5]*Nanomaterials Research Institute, Kanazawa University, Kanazawa, Japan*

[6]*Department of Physics, Tohoku University, Sendai, Japan*

[7]*Department of Applied Physics, University of Tokyo, Tokyo, Japan*

[8]*Institute for Quantum Matter and Department of Physics and Astronomy, Johns Hopkins University, Baltimore, MD, USA*

[9]*Trans-scale Quantum Science Institute, University of Tokyo, Tokyo, Japan*

---

[*] These three authors contributed equally

**The recent discoveries of strikingly large zero-field Hall and Nernst effects in antiferromagnets $Mn_3X$ ($X$ = Sn, Ge) have brought the study of magnetic topological states to the forefront of condensed matter research and technological innovation. These effects are considered fingerprints of Weyl nodes residing near the Fermi energy, promoting $Mn_3X$ ($X$ = Sn, Ge) as a fascinating platform to explore the elusive magnetic Weyl fermions. In this review, we provide recent updates on the insights drawn from experimental and theoretical studies of $Mn_3X$ ($X$ = Sn, Ge) by combining previous reports with our new, comprehensive set of transport measurements of high-quality $Mn_3Sn$ and $Mn_3Ge$ single crystals. In particular, we report magnetotransport signatures specific to chiral anomalies in $Mn_3Ge$ and planar Hall effect in $Mn_3Sn$, which have not yet been found in earlier studies. The results summarized here indicate the essential role of magnetic Weyl fermions in producing the large transverse responses in the absence of magnetization.**

**Introduction**

In the field of quantum materials, topology — a mathematical concept that describes the robustness of form — has inspired a new era by introducing novel topological phases of matter[1-3]. Until recently, most of the associated discoveries take place in weakly-interacting materials in which electron correlation effects have only a minor role. Now the hunt for new topological phases is shifting toward strongly correlated electron systems. The notion of strong electronic correlation provides a common thread linking a wide variety of fascinating emergent phenomena in quantum materials, ranging from high-$T_c$ superconductivity to fractional quantum Hall effect (FQHE). Its convergence with topology affords an exceptional venue to explore topological order in magnetism. A key discovery along this direction is the surprisingly large anomalous Hall effect (AHE) in chiral antiferromagnets[4-6], which has triggered extensive research efforts in various fields, ranging from topological condensed matter physics, spintronics, and energy harvesting technology[7-16]. A serious challenge facing condensed-matter and material physicists is revolutionizing information technology that breaks through the fundamental limit of Moore's law, namely, developing strategies that enhance the processing speed and energy efficiency simultaneously. With this regard, spintronics has provided cutting-edge technologies, introducing non-volatile forms of logic and memory devices. Traditionally focused on ferromagnetic materials, recent studies have revealed further benefits using antiferromagnets: antiferromagnets possess ultrafast dynamics and are insensitive to stray fields that perturb neighboring cells, allowing them to exceed the capacity of the ferromagnetic counterparts[17, 18]. Thus, the large AHE in antiferromagnets observed at room temperature may serve as a readout for a spintronic device[1, 16-18]. Another remarkable discovery is the substantial anomalous Nernst effect (ANE) accompanying the large AHE without net magnetization[9, 13, 15, 19]. This feature indicates that the control of Berry curvature is essential for enhancing the transverse thermoelectric effect, which has inspired new development in thermoelectric converters[9, 20-23].

Understanding of the mechanism behind the unusually large AHE and ANE in antiferromagnets serve as the basis for their application in modern technologies; this is where the novel topological quasiparticles — Weyl fermions — comes into play. In a metallic or semimetallic material, the double electronic band degeneracy enforced by the Kramers theorem is lifted with the breaking of either time-reversal or spatial inversion symmetry. The linearly dispersing touching point of two non-degenerate bands is known as Weyl nodes, around which the Weyl Hamiltonian $H_W = \pm \hbar v_F \boldsymbol{k} \cdot \sigma$ describes the electronic states[24, 25]. According to the dimensionality of the electronic band touching, the topological states can be classified into topological insulators, nodal points, and nodal lines; the Weyl node is a typical example of the nodal points. In this case, the emergent quasiparticle represents a condensed-matter physics analog of the long-anticipated relativistic Weyl fermion that remains unobserved in high-energy physics experiments. The Weyl nodes carry quantized magnetic charges that act as a source or drain of Berry curvature[24]. The Berry curvature can be regarded as a fictitious magnetic field in momentum space, as it behaves in the same manner as a real magnetic field under symmetry operations[26, 27]. In contrast with Dirac fermions in graphene, the Weyl node is a three-dimensional object; namely, the corresponding Weyl Hamiltonian involves all three Pauli matrices that form the basis of the Hilbert space. As a result, any perturbation in the form of a linear superposition of the Pauli matrices cannot annihilate the Weyl points but merely shift its position in momentum space. The magnetic charge is protected as in the Gauss's law for an electronic charge in real space. This topological protection of Weyl nodes against perturbation renders them appealing potential sources of real-life applications.

The Weyl semimetal (WSM) state occurs in two situations, requiring either a broken time-reversal symmetry (TRS) or a broken spatial inversion symmetry (IS); the latter has been established in weakly-interacting electron systems with spectroscopic evidence of Weyl fermions. The experimental quest for a TRS-breaking WSM is far more challenging, yet it features advantageous properties from both fundamental and practical standpoints. The TRS-breaking or magnetic WSM provides the possibility of realizing a model system that exhibits macroscopic spontaneous responses due to Weyl fermions in zero field, which are absent in the IS-breaking WSM. For instance, the substantial net Berry curvature inherent to the Weyl nodes, combined with the magnetism, may yield unusually large anomalous transport and optical properties in TRS-breaking WSMs. Moreover, the magnetic texture provides a handle for manipulating the Weyl fermions[4, 5, 8, 28], making the magnetic WSMs particularly attractive in spintronic applications[16].

In noncollinear antiferromagnets, the combination of the symmetry of the spin structure and the underlying band topology allows for the realization of a TRS-breaking WSM[8, 28, 29]. When the Weyl nodes reside close to the Fermi energy, they may manifest through large intrinsic AHE and ANE due to the significant enhancement in the Berry curvature. In fact, in addition to the chiral anomaly in magnetotransport, the observation of AHE and ANE in antiferromagnets have provided first crucial experimental signatures of magnetic Weyl fermions[8]. Here, we provide an overview of transport measurements in chiral antiferromagnet $Mn_3X$ ($X$ = Sn, Ge) that features unusually large AHE and ANE. Combined with spectroscopic studies and first principles calculations, the comprehensive set of transport signatures in $Mn_3X$ ($X$ = Sn, Ge) serves as an effective probe of magnetic Weyl fermions, paving an important avenue for future identification of the Weyl semimetal state in strongly correlated materials.

**Basic properties of $Mn_3X$ ($X$ = Sn, Ge)**

Kagome-based metals have recently become a new focus in the field of quantum materials for their topological electronic structures, as well as large electromagnetic and transport responses[4-6, 9, 11, 12, 15, 23, 30, 31]. $Mn_3Sn$ and $Mn_3Ge$ are hexagonal chiral antiferromagnets (space group P6$_3$/*mmc*) that comprise breathing-type kagome layers of Mn atoms in the *ab*-plane (Fig. 1a); these kagome planes are stacked along the *c*-axis[32]. Neutron diffraction studies in the magnetically ordered phase revealed an inverse triangular spin structure, in which Mn moments are aligned within the kagome layers, forming, a 120 ° ordering with a negative vector chirality and a $q = 0$ ordering wave vector; the ordered moment is about 3 $\mu_B$/Mn ($\mu_B$, Bohr magneton) in $Mn_3Sn$ and about 2.5 $\mu_B$/Mn in $Mn_3Ge$[33-36]. This chiral spin structure reduces the hexagonal crystal structure to orthorhombic in both compounds. Within each triangular motif, only one of the three Mn moments points along the local easy-axis ($[2\bar{1}\bar{1}0]$ for $Mn_3Sn$ (Fig. 1a) and $[01\bar{1}0]$ for $Mn_3Ge$ (Fig. 1b)). Small canting of the remaining two spins in the *ab*-plane induces weak ferromagnetism with a tiny net moment of about 3 m$\mu_B$/Mn in $Mn_3Sn$ and about 7 m$\mu_B$/Mn in $Mn_3Ge$[33, 35]. The second-order nature of the magnetic transition suggests that the spin structure of $Mn_3X$ ($X$ = Sn, Ge) belongs to the $D_{6h}$ point group, which allows four distinct 120 ° spin structures with $q = 0$. Recent polarized neutron diffraction studies on single crystals constrain the ground-state spin structure to $E_{1g}$-symmetry type, for which the weak ferromagnetism is inevitable[37, 38]. This information is essential for verifying the theoretical proposals of Weyl nodes based on band structure calculations[8, 28, 39, 40]. Owing to the competition between the single-ion anisotropy and the Dzyaloshinskii-Moriya (DM) interaction, the in-plane anisotropic energy of Mn is absent up to fourth order in $Mn_3X$ ($X$ = Sn, Ge), and six-order anisotropy is necessary for selecting a unique ground-state spin structure[33, 34, 41-43]. With such small anisotropic energy, the coupling between a magnetic field and the weak ferromagnetic moment takes on a key role that enables magnetic-field control of the antiferromagnetic spin structure.

The Mn moments residing on an octahedron made of two triangles of neighboring kagome bilayers constitute a cluster magnetic octupole; the inverse triangular spin structure can be viewed as a ferroic order of cluster magnetic octupoles (Fig. 1b)[7, 44, 45]. $Mn_3Sn$ serves as a textbook example for the developing cluster multiple (CMP) theory, which identifies the symmetry equivalence of noncolinear antiferromagnetic (AFM) spin structures with ferromagnetic (FM) states by considering CMP as the order parameter of noncollinear AFM orders[44, 45]. The large magneto-optical Kerr effect (MOKE) has been discovered in Mn$_3$Sn as the first example in the antiferromagnetic metal and enables visualization of the magnetic octupole domains and their switching under an applied magnetic field[7]. The interplay between CMP in real-space magnetic domains and the Weyl points in momentum space may yield intriguing topological electric transport that remains to be explored[46].

Below the Néel temperature $T_N \approx 430$ K, Mn$_3$Sn enters the inverse triangular spin state, which is taken over by helical ordering and a cluster glass phase at low temperatures[4, 9, 34, 43, 47-49], depending on the Mn concentration. In contrast, Mn$_3$Ge retains the inverse triangular spin structure down to at least 0.3 K, as is evident from the magnetization $M$ curves of our newly synthesized Mn$_3$Ge single crystals (Fig. 2e): $M$ undergoes a sharp transition at the Néel temperature $T_N \approx 372$ K, and no additional transition occurs down to 0.3 K. The in-plane magnetization reaches a saturation value of about 7 m$\mu_B$/Mn at $B = 0.1$ T and $T < 100$ K, which is approximately ten times greater than the magnetization for $B \parallel [0001]$.

The hexagonal structure of bulk Mn$_3$Sn and Mn$_3$Ge are stable only in excess Mn[4, 5, 33, 35], leading to Mn-substituted Sn/Ge sites. The off-stoichiometry is reflected in a short electronic mean free path of the order several nm at room temperature, comparable to the lattice parameters[13, 19]. This compositional disorder may alter the magnetic and anomalous transport properties of the material and deserves detailed studies. We have recently investigated the behavior of Si-doped Mn$_3$Ge crystals, and find that Si doping into the Ge sites provides a means for systematically examining how the physical properties evolve with excess Mn. By doping Si into Ge sites, the level of extra Mn $x$ is enhanced concomitantly; the lattice parameters and the unit cell volume exhibit a slight linear decrease with increasing amount of extra Mn. The volume reduction rate $dV/dx$ (= -39 ~ -48 Å$^3$/Mn) in Mn$_{3+x}$(GeSi)$_{1-x}$ is nearly identical to the value found in Mn$_{3+x}$Ge$_{1-x}$, implying that Si acts merely as an impurity ion at the Ge site. The magnetization of Si-doped Mn$_3$Ge samples shows temperature variation similar to that of undoped Mn$_3$Ge, while the Néel temperature decreases monotonically with increasing amount of Si doping (Fig. 2e inset).

## 2. Observation of Anomalous Transport in Mn$_3$X (X = Sn, Ge)

**Anomalous Hall effect (AHE).** The AHE is traditionally considered to be proportional to net magnetization, and therefore is expected to arise only in ferromagnetic materials[50]. In the past decade, theoretical advances in understanding the mechanism of AHE has established that the intrinsic contribution to the anomalous Hall conductivity (AHC) is the sum of the Berry curvature of all the occupied bands[26, 27, 51], such that

$$\sigma_{xy} = \frac{e^2}{\hbar} \int_{BZ} \frac{d\mathbf{k}}{(2\pi)^3} \Omega_{n,z}(\mathbf{k}) f_{n,k} , \qquad (1)$$

where $\Omega_{n,z}(\mathbf{k})$ is the out-of-plane component of the Berry curvature with the band index *n* and $f_{n,k}$ is the Fermi-Dirac distribution function. Namely, the intrinsic AHE is driven not by the magnetization but by the Berry curvature determined entirely by the band topology of a perfect crystal. That is to say, the sizable AHE can be realized even in the absence of net magnetization, such as in spin liquids and antiferromagnets, as long as the system sustains a substantial net Berry curvature in the momentum space[24, 26-28, 39, 44, 52]. In fact, the first experimental detection of this kind was made in 2009 in the spin liquid compound Pr$_2$Ir$_2$O$_7$, where a large zero-field Hall conductivity of about 10 $\Omega^{-1}$cm$^{-1}$ is observed in the absence of magnetization in a spin-ice state[29, 53, 54]. Systems harboring Weyl points offer an ideal physical setting that promotes strongly enhanced AHE without net magnetization, owing to the divergent Berry curvature in the immediate vicinity of the Weyl points. Following the observation of AHE in $Pr_2Ir_2O_7$, the seminal paper by Wan *et al*[24] proposed that TRS-breaking Weyl nodes can be stabilized by the all-in all-out magnetic order, another type of magnetic cluster octupole, in the family of pyrochlore iridates $R_2Ir_2O_7$, ($R$ = Y and lanthanides) which brought the concept of magnetic WSM to the center stage of condensed matter physics.

In 2015, the first observation of the large AHE in an antiferromagnet was achieved using the chiral AFMs $Mn_3X$ ($X$ = Sn, Ge)[4, 5]. At room temperature, the Hall resistivity $\rho_H$ of $Mn_3Sn$ measured under in-plane magnetic fields $B \parallel [2\bar{1}\bar{1}0]$ and $B \parallel [01\bar{1}0]$ displays a sharp hysteresis loop with a small coercivity of approximately 100 Oe (Fig. 2a)[4]. The magnetic-field dependence of $\rho_H$ is anisotropic; the sweep of the in-plane fields generates a substantial zero-field Hall component and a narrow hysteresis in $\rho_{yz}$ and $\rho_{zx}$, whereas the out-of-plane field $B \parallel [0001]$ yields a linear-in-$B$ response in $\rho_{xy}$ without hysteresis. The in-plane-field responses, $\rho_{yz}$ and $\rho_{zx}$, are unusually large for an antiferromagnetic material and are comparable to the values found in strong ferromagnets[26]. By contrast, the magnetization isotherms $M$ ($B$) observed under the in-plane magnetic fields show a vanishingly small spontaneous (zero-field) magnetization $M$ of about 3 m$\mu_B$/Mn originated from spin canting (Fig. 2b)[4]. The similar hysteretic and anisotropic behavior of $\rho_H(B)$ and $M$ ($B$) indicates that the rotation of antiferromagnetic domains leads to the sign change of the Hall effect. Figure 2c depicts the temperature dependence of the zero-field Hall conductivity, $\sigma_{yz}$, $\sigma_{zx}$, and $\sigma_{xy}$, obtained under $B \parallel [2\bar{1}\bar{1}0]$, $B \parallel [01\bar{1}0]$, and $B \parallel [0001]$, respectively[4]. The zero-field values of $\sigma_{yz}$ and $\sigma_{zx}$ gradually increase from the room-temperature value of about 20 $\Omega^{-1}cm^{-1}$ on cooling, reaching a maximum value of about $130\ \Omega^{-1}\text{cm}^{-1}$ for $-\sigma_{zx}$ and about $70\ \Omega^{-1}\text{cm}^{-1}$ for $-\sigma_{yz}$ before dropping sharply upon entering the low temperature spin-glass phase at $T < 50$ K. In contrast, $\sigma_{xy}$ becomes finite only in the low temperature spin-glass phase and its magnitude increases strongly on cooling for $T < 50$ K (Fig. 2c)[4].

The negligible $\sigma_{xy}$ at high temperatures is in line with the in-plane polarization of the magnetic octupole due to coplanar 120 ° spin structure, whereas the drastic increase in $\sigma_{xy}$ observed at $T < 50$ K may involve a topological Hall effect[4]. For $T < 20$ K, the Hall measurements in polycrystalline Mn₃Sn samples reveal sharp peaks of $\rho_{xy}(B)$ at about $\pm 0.8$ T, suggesting a topological Hall contribution $\rho_{xy}^T$ that dominates the Hall signal in the low-$T$ regime[55]. The spontaneous magnetization at $T < 20$ K increases by over an order of magnitude relative to the high-temperature value; the coercivity also rises dramatically from only a few hundred oersteds at high temperatures to about 1 T at 2 K. Similar low-$T$ behavior in both $\rho_{xy}$ and $M$ is observed in polycrystals of a sister material Mn₃Ga[56]. The topological Hall effect arises from scalar chirality subtended by three neighboring spins of a non-coplanar spin texture[26, 53, 54, 57-59]. Recent studies have highlighted the topological Hall effect as a transport signature of a magnetic skyrmion state, as have been seen in various noncentrosymmetric skyrmion-hosting materials MnSi, FeGe and MnGe[60-62]. Therefore, the above experiments indicate that Mn₃Sn may undergo a transition from the high-$T$ coplanar 120 ° spin structure with zero scalar spin chirality to a low-$T$ noncoplanar spin ordering, in which the sizeable scalar spin charily acts as a fictitious magnetic field in real space, thereby inducing nontrivial transport responses. Given the strongly anisotropic behavior of AHE and $M$ observed in Mn₃$X$ systems, detailed single-crystal studies are necessary for an in-depth exploration of the topological Hall effect and the magnetic skyrmion scenario.

In contrast with Mn₃Sn, we show that the Hall conductivity $\sigma_{yz}$ of our Mn₃Ge single crystals increases on cooling and saturates at approximately 300 $\Omega^{-1}\text{cm}^{-1}$ below about 100 K (Fig. 2f) owing to the absence of additional magnetic phase transition at low temperatures[5, 6]. Contrary to the ferromagnetic case, neither of $\sigma_{yz}$ nor $\sigma_{zx}$ follows the behavior of *M(T)* (Figs. 2e, f), which are consistent with earlier reports[5]. With a slight increase in the extra Mn concentration, the saturated value of the AHC significantly declines (Fig. 2f), suggesting that the excess Mn controls the number of the conduction electrons and thereby tunes the Fermi level $E_F$, as the electron states at $E_F$ are constituted solely by the 3*d* electrons[8, 28].

**Anomalous Nernst effect (ANE).** The ANE is another key quantity that measures the Berry curvature. Unlike the AHE that involves the Berry curvature over all occupied bands, the ANE is determined by the Berry curvature lying close to the Fermi energy $E_F$[9, 27, 63]. The Nernst signal $S_{ji}$ is obtained using the same electrodes as in the Hall effect measurements, by applying temperature gradient instead of charge current; namely, an induced transverse voltage $V_j$ perpendicular to both the magnetic field *B* and the heat current $Q_i \propto -\nabla_i T$ is measured at various temperatures. The Nernst coefficient $S_{xy} = E_y/(-\nabla_x T)$ is related to the anomalous transverse thermoelectric conductivity $\alpha_{xy}$ such that $S_{xy} = \alpha_{xy}\rho_{xx} - \rho_{xx}\sigma_{xy}S_{yy}$, where $\rho_{xx}$, $\sigma_{xy}$, and $S_{yy}$ are the longitudinal resistivity, the Hall conductivity, and the Seebeck coefficient, respectively[4]. The connection of $\alpha_{xy}$ with the Berry curvature near the Fermi energy is well established, namely,

$$\alpha_{xy} = \frac{e}{Th}\int_{BZ}\frac{d\mathbf{k}}{(2\pi)^3}\Omega_{n,z}(\mathbf{k})\{(\epsilon_{n,\mathbf{k}} - \mu)f_{n,\mathbf{k}} + k_BT\log[1 + e^{-\beta(\epsilon_{n,\mathbf{k}}-\mu)}]\} \quad (2)$$

where $\Omega_{n,z}(\mathbf{k})$ is the out-of-plane component of the Berry curvature, $\epsilon_{n,\mathbf{k}}$ is the band energy, $\mu$ is the chemical potential, and $f_{n,\mathbf{k}}$ is the Fermi-Dirac distribution function, with *n* the band index[63].

Figure 2d shows the key findings of the substantial room-temperature ANE in Mn$_3$Sn reported by Ikhlas *et al.* (2017)[9]. Similar to the Hall conductivity $\rho_H$ and the magnetization *M*, the field dependence of $-S_{ji}$ is strongly anisotropic; the in-plane field components $-S_{yz}$ and $-S_{zx}$ exhibit sharp hysteretic jump with a substantial zero-field component of about 0.3 µVK$^{-1}$, more than three orders of magnitude higher than the expected value according to the scaling relation for ferromagnetic materials (Fig. 6b), while $-S_{yx}$ remains zero within experimental resolution without showing any hysteresis. The comparison between the field dependence of $-S_{ji}$ and *M* help unveil the driving mechanism behind the ANE. Although $-S_{yz}$ and *M* measured under $B \parallel [2\bar{1}\bar{1}0]$ display nearly identical hysteresis loops in the low-field regime, $-S_{yz}$ levels off at higher fields rather than following the linear increase of *M*. Given that the *B*-linear component of *M* arises from the field-induced spin canting, the distinction between field dependence of ANE and *M* indicates that the ANE is insensitive to the spin canting, but instead is governed by the Berry curvature tied in with the order parameter, namely, the cluster magnetic octupole[44, 45]. The temperature dependence of the in-plane zero-field components, $-S_{yz}$ and $-S_{zx}$, peak at around 200 K with a maximum value of about 0.6 µVK$^{-1}$ for Mn$_{3.06}$Sn$_{0.94}$. With a slightly more Mn in Mn$_{3.09}$Sn$_{0.94}$, the peak in $-S_{yz}$ and $-S_{zx}$ shifts to a higher temperature of about 250 K with a suppressed maximum value of about 0.3 µVK$^{-1}$. The dramatic change in ANE with only 1% of variation in Mn concentration indicates the proximity of the Fermi energy to the Weyl points, which is discussed in more detail in the next section.

Similar to the case of Mn$_3$Sn, our newly obtained Nernst coefficient $-S_{yz}$ of Mn$_3$Ge single crystals shows a step-like profile with a sizeable zero-field component at room temperature, as shown in Fig. 2g. On cooling, the zero-field components of $-S_{yz}$ and $-S_{zx}$ behave similarly and reach a maximum of about 1.35 μVK$^{-1}$ near 100K (Fig. 2h), which is more than twice larger than the value reported for Mn$_3$Sn[9]. The recent studies by Wuttke *et al* (2019)[15] also find that the temperature dependence of zero-field $-S_{yz}$ and $-S_{zx}$ are nearly identical, with magnitude consistent with our finding. Under a high magnetic field of 14 T, $-S_{yz}$ and $-S_{zx}$ are further enhanced to about 1.5 μVK$^{-1}$ while the broad peak remains at around 100 K (Fig. 4a). The out-of-plane field component $-S_{xy}$ is negligibly small compared to $-S_{yz}$ and $-S_{zx}$ in the studied temperature range. The behavior of $S_{ji}$ observed in Mn$_3$X (X = Sn and Ge) share remarkable similarities with the low-temperature field induced ANE of prototypical nonmagnetic Weyl semimetals TaAs and TaP, in which the peak position of $S_{ji}$ provides a rough estimate of the lowest Weyl node energy relative to $E_F$[64]. For $T > 50$ K, the ANE observed in TaAs and TaP gives way to the *B*-linear ordinary contribution[64], whereas the ANE in Mn$_3$X (X = Sn and Ge) dominates the Nernst signal over a much wider temperature range up to room temperature.

To comprehensively characterize the thermoelectric properties, we have newly measured the longitudinal Seebeck coefficient, $S_{ii}(T)$, of Mn$_{3.04}$Ge$_{0.96}$ for $Q \parallel [0001]$ and $Q \parallel [01\bar{1}0]$ configurations (Fig. 2h inset). The behavior of $S_{ii}(T)$ is again strongly anisotropic; while $S_{zz}$ remains negative in the entire measured temperature range and is weakly dependent on temperature, $S_{yy}$ exhibits a sign change from positive to negative below about 200 K and peaks negatively at about 60 K. The sign change is presumably related to the phonon drag effect[65], as found in the case of Mn$_3$Sn[9,19].

**Anomalous thermal Hall effect (ATHE).** As the thermal counterpart of the AHE, the anomalous thermal Hall conductivity, $\kappa_{ij}^A$, provides essential clues about the driving mechanism behind the dramatically enhanced room-temperature AHE in Mn$_3$X (X = Sn, Ge). Similar to the behavior of the Hall conductivity and the Nernst signal, the thermal Hall conductivities, $\kappa_{xz}(B)$ and $\kappa_{zy}(B)$, in Mn$_3$Sn show a sharp hysteretic jump corresponding to the anomalous component[19, 66] (Figs. 3a, b). The anomalous Wiedemann-Franz (WF) law is defined as

$$L_{ij}^A = \frac{\kappa_{ij}^A}{T\sigma_{ij}^A} = L_0 = \frac{\pi^2}{3}\left(\frac{k_B}{e}\right)^2, \qquad (3)$$

which relates the ratio between anomalous thermal and electrical Hall conductivities with the Sommerfeld number $L_0$. In conventional ferromagnets such as Fe, Ni, and NiCu alloys, the magnitude of $L_{ij}^A$ displays a downward deviation from $L_0$ in the high temperature regime due to inelastic scattering, and approaches $L_0$ only in the zero-temperature limit[19, 65-68] (Fig. 3c). In Mn$_3$Sn, however, the $L_{ij}^A$ is nearly temperature independent and satisfies the WF law in the entire studied temperature range from room temperature down to about 50 K[19, 66, 69] (Fig. 3c). This behavior suggests a negligible effect of inelastic scattering on the anomalous transverse transport of Mn$_3$Sn, excluding phonon contributions or skew scattering by magnetic excitations as a possible cause of the room-temperature AHE and ATHE[68, 70]. The validity of the WF law over a wide temperature range in Mn$_3$Sn further supports the intrinsic Berry curvature mechanism for its large anomalous transverse effects[4, 9, 26, 56, 63, 66, 71].

In contrast, the $L_{zx}^A$ of Mn$_3$Ge follows the anomalous WF law $L_{zx}^A \approx L_0$ below 100 K down to 0.3 K but deviates from $L_0$ above 100 K[13] (Fig. 3d). The high-temperature violation of the anomalous WF law in Mn$_3$Ge is unlikely to result from inelastic scattering, but instead may come from a strong energy distribution of the Berry curvature near $E_F$ that results in a mismatch between its electrical and thermal summations over the Fermi surface. The difference in the valid $T$ range of the anomalous WF law suggests that the Weyl points play a vital role in Mn$_3$Sn in shaping the behavior of observed temperature dependence of $L_{zx}^A$. As for Mn$_3$Ge, aside from the contribution from Weyl points, its Berry spectrum near $E_F$ is influenced by an anti-crossing gap induced by spin-orbit coupling (SOC), which serves as the primary source of the downward deviation in $L_{zx}^A$ above 100 K[13].

**Evidence for Magnetic Weyl Fermions**

Following the experimental discoveries of the large AHE and ANE, Mn$_3$X ($X$ = Sn, Ge) is proposed as the magnetic version of the Weyl semimetal (Weyl magnet) based on first principle calculations[8, 28]. The coplanar 120° spin structure of Mn$_3$Sn and Mn$_3$Ge (Figs. 1a and b) lowers the original hexagonal crystal symmetry to orthorhombic, which possesses a nonsymmorphic symmetry and two mirror reflections, $M_x T$ and $M_y T$, with time reversal $T$ added. These symmetry properties set strong constraints on the positions of the Weyl points. That is, the mirror symmetries ensure that the Weyl points appear along a $k$ direction parallel to the local easy magnetization axis, namely, the $x$-axis for Mn$_3$Sn and $y$-axis for Mn$_3$Ge; the corresponding $k$-space location of Weyl points are shown in Figs. 1d and 1e, respectively[7, 8]. By applying a rotating magnetic field in the $x$-$y$ plane, one may shift the locations of the Weyl points governed by the underlying spin texture along a hypothetical nodal ring obtained in the absence of the spin-orbit coupling, as illustrated in Fig. 1d.

The calculated band dispersion in both Mn$_3$Sn and Mn$_3$Ge reveals multiple pairs of type-II Weyl nodes at various energies[28], which possess finite density of state at the nodal point owing to the touching between electron and hole pockets. Among those, the most relevant to transport properties are the pairs sitting close to the Fermi level $E_F$. In the case of Mn$_3$Sn, the two pairs of Weyl points lying in the vicinity of $E_F$ arise from electron-hole band crossings along the K-M-K direction[8], as shown in Fig. 1c. Each pair consists of two Weyl points of opposite chirality, with $W_1^+$ and $W_2^-$ lying at $E \sim E_F + 60$ meV and their partners $W_1^-$ and $W_2^+$ appearing at a slightly higher energy $E \sim E_F + 90$ meV. The experimental quest for the Weyl fermions involves various experimental probes combined with theoretical analyses, as presented below. In this section, we summarize the evidence of Weyl points obtained for Mn$_3$Sn and Mn$_3$Ge and present our updated analyses of the transverse thermoelectric coefficient, $\alpha_{ji}$, observed in Mn$_3$Ge based on the density functional theory (DFT) calculations.

**Analysis of the anomalous transverse coefficients and carrier doping effect.** We begin by discussing the temperature variation of the anomalous transverse thermoelectric conductivity[72], $\alpha_{xz} = (\rho_{xx}S_{xz} - \rho_{xz}S_{zz})(\rho_{xx}\rho_{zz} + \rho_{xz}^2)$. The $-\alpha_{xz}$ observed in Mn$_3$Sn and Mn$_3$Ge behaves similarly, namely, $-\alpha_{xz}$ gradually increases with decreasing temperature, forms a broad maximum, and then declines at lower temperatures (Figs. 4b, h, i)[15]. In both materials, the anomalous Nernst signal $S_{xz}$ dominates $\alpha_{xz}$ (Fig. 4c)[15]. The analysis presented in Wuttke *et al.* (2019)[15] is based on a minimal model describing a single pair of Weyl points near $E_F$ formed by a crossing of linearized bands. The resulting fitting formula for $-\alpha_{xz}$ involves three key parameters, that is, the Berry curvature strength, $\widetilde{\Omega}$, near $E_F$, the **k**-space nodal separation, $\Delta k$, between the two Weyl points, and the Weyl point energy, $E$, relative to $E_F$. For a system with multiple pairs of Weyl points, the $E$ value provides an estimate of the lowest possible Weyl point energy. For both Mn$_3$Sn and Mn$_3$Ge, such a model well describes the experimental $-\alpha_{xz}$ (solid lines in Figs. 4b, d)[15], and the $E$ value of Weyl points yielded by the best fit is in reasonable agreement with the predicted energy given by band structure calculations[8, 28].

We next present a comparison between the experimentally observed $-\alpha_{ji}$ in $Mn_3Sn$ and $Mn_3Ge$ with our DFT calculations. The $Mn_3Sn$ and $Mn_3Ge$ samples usually contain extra Mn, as mentioned earlier. Since the states at the $E_F$ is occupied solely by 3d-electrons, a small amount of Mn doping may generate a sizeable shift of the chemical potential relative to the $E_F$ of the stoichiometric case. A 1% Mn-doping in $Mn_3Sn$ may shift up the chemical potential by about 6 meV from $E_F$; thus, the two Mn compositions, $Mn_{3.06}Sn_{0.94}$ and $Mn_{3.09}Sn_{0.91}$, have the chemical potential of $E = E_F + 36$ meV and $E = E_F + 54$ meV, respectively, as confirmed by ARPES measurements[7]. Moreover, ARPES study finds a significant bandwidth renormalization and quasi-particle damping. Thus, in the calculation of the anomalous transverse thermoelectric conductivity, we take both the real and imaginary parts of the self-energy into account, such that

$$\Sigma = \left(1 - \frac{1}{Z}\right)\epsilon - iA\epsilon^2 - i\gamma \qquad (4)$$

The Z value is estimated to be 0.2 for $Mn_3Sn$ according to the renormalization factor estimated by the ARPES and specific heat measurement[7, 8]. Given the larger bandwidth of $Mn_3Ge$ in comparison with $Mn_3Sn$[44, 45], a larger Z value of 0.5 is chosen. The coefficient $1/ZA$ determines the energy window within which the quasi-particle is well defined. We assume a window size of 40 (100) meV for $Mn_3Sn$ ($Mn_3Ge$) based on the ARPES measurement[7] of $Mn_3Sn$ so that the size covers the energy scale for Weyl fermions relevant for both compounds. Thus, $AZ^2 = 5$ eV$^{-1}$ for both $Mn_3Sn$ and $Mn_3Ge$. The solid curves in Figure 4h are the calculated $-\alpha_{zx}/T$ for $Mn_{3.06}Sn_{0.94}$ and $Mn_{3.09}Sn_{0.91}$ that are dominated by the Berry curvature stemming from the low-energy Weyl points. They reproduce the overall temperature dependence of the experimental $-\alpha_{zx}/T$ yet overestimate the magnitude by nearly a factor of 2. Note that we limit the temperature range for the comparison of theory and experiment up to 200 K; the temperature variation of the magnetization $M$ is negligible in this $T$ regime, and thus the spin fluctuations should have only a minor effect on the transport properties.

In the case of Mn$_3$Ge, our DFT calculations predict the presence of several Weyl points near $E - E_\mathrm{F} = 78$ meV (Fig. 4f) and 63 meV (Fig. 4g) under $B \parallel [01\bar{1}0]$ on the $k_z = 0$ and $\pm 0.137$ (Å$^{-1}$) planes, respectively, where the energy gap between the electron and hole bands becomes vanishingly small within the Brillouin zone (BZ). These Weyl points are located at non-high-symmetric positions in the BZ and harbor type–II Weyl fermions; their energies and positions in the $\boldsymbol{k}$-space are in good agreement with the W$_1$ and W$_3$ pairs reported in Yang $et\ al$ (2017)[28]. The theoretical calculation suggests that a 1% Mn-doping may shift the chemical potential up by about 17 meV in Mn$_3$Ge, and, as a result, the experimental Hall conductivity, $\sigma_{ji}$, shown in Fig. 2f decreases from 300 $\Omega^{-1}\mathrm{cm}^{-1}$ in Mn$_{3.03}$Ge$_{0.97}$ to 170 $\Omega^{-1}\mathrm{cm}^{-1}$ in Mn$_{3.04}$Ge$_{0.96}$. The calculated energy spectrum of $\sigma_{ji}$ undergoes such a sharp change at $E - E_\mathrm{F} \approx 60$ meV; correspondingly, the calculated $-\sigma_{ji}$ and $-\alpha_{ji}$ with the chemical potential of $E - E_\mathrm{F} = +50$ meV and $E - E_\mathrm{F} = +68$ meV roughly reproduces the $T$ dependence and magnitudes observed for Mn$_{3.03}$Ge$_{0.97}$ and Mn$_{3.04}$Ge$_{0.96}$ (Fig. 4h).

**Spectroscopic investigations of the electronic band structure.** Identification of the band structure and the locations of Weyl points in the momentum space provides direct evidence for the Weyl fermions. In weakly-interacting electron systems with broken inversion symmetry, clear spectroscopic evidence of Weyl points and surface Fermi arcs has been established[1, 73]. The search of Weyl fermions in magnetic materials that breaks time-reversal symmetry is far more challenging because strong correlation effects due to magnetism inevitably lead to diffusive character of the Weyl excitations and prevent their detection by spectroscopic probes. While recent angle-resolved photoemission spectroscopy (ARPES) experiments find some evidence for magnetic Weyl fermions in ferromagnetic Co-based materials[74, 75], the present Mn$_3$X (X = Sn, Ge) compounds feature much stronger electron correlations than the Co-based systems, as reflected by the large renormalization factor of five. Such dramatic bandwidth renormalization is comparable to that of high-temperature superconductors and heavy fermion compounds[76], therefore posing an intrinsic challenge for direct observation of Weyl fermions. Figures 5 a, b, and c summarize the ARPES spectra of Mn$_3$Sn reported by Kuroda *et al*. (2017)[8]. The $k_z$-dispersion of ARPES intensity along the H-K-H high symmetry direction reveals a quasiparticle peak near $E_F$ at a photon energy of $hv = 103$ eV (the blue arrow in Fig. 5a). This feature indicates that $hv = 103$ eV selectively measures the $k_z = 0$ band dispersion involving the low-energy Weyl points. The ARPES constant-energy surface at $E_F$ obtained using $hv = 103$ eV captures six elliptical contours on the $k_x - k_y$ plane, in qualitative agreement with the topological features of the electron-type Fermi pocket predicted by the DFT calculation (Fig. 5b). More importantly, this Fermi pocket is derived from the electron band that generates the Weyl points at its crossing with a hole band at $E \sim E_F + 60$ meV. The evolution of this electron band is traced by the $E$-$k_x$ cuts measured at various $k_y$ along the K-M-K line (Fig. 5c), consistent with the theoretical band dispersion after renormalizing the energy scale by a factor of five. That is, the electron band (the red line) approaches the hole band (the blue line) with increasing $k_y$, form the Weyl nodes, and again moves apart from the hole band. The momentum distribution curve (MDC) measured at 60 K displays two additional anomalies only for the $k_x$-cut made exactly along K-M-K, corresponding to the linear band crossing points between the electron and hole bands (Fig. 5c). Although the strong correlation effects hinder the observation of direct signatures of Weyl points and surface Fermi arcs, these findings in APRES measurements are essential, as they identify the two bands that form the low-energy pair of Weyl nodes (Figs. 1c, d) and verify the calculated band structure.

**Chiral anomaly in Mn$_3$X (X = Sn, Ge).** The gapless nature of the Weyl fermions renders their acute responses in low-energy probes, with the hallmark feature in magnetotransport known as the chiral anomaly. In the presence of a high magnetic field, the Weyl crossings are quantized into Landau levels (LLs), with a chiral lowest LL for each Weyl node. Namely, electrons occupying the lowest LL are divided into left- and right-moving groups of opposite chirality. Hypothetically, these two groups do not mix, leading to separate number conservation of the three-dimensional left- and right-handed Weyl fermions. Once an electric field $E \parallel B$ is applied, the conservation of chirality is no longer valid, as a result of the mixing between the left-and-right movers[77, 78]. This phenomenon is known as the chiral anomaly. In such a case, charge pumping between the Weyl nodes of opposite chirality generates the positive longitudinal magnetoconductivity (LMC) or the negative magnetoresistivity (LMR) only for the $E \parallel B$ configuration. Meanwhile, the absence of the chiral anomaly for the perpendicular configuration $E \perp B$ leads to a negative longitudinal magnetoconductivity (LMC) or a positive magnetoresistivity (LMR) due to the ordinary scattering mechanism. Thus, the combination of the positive LMC for $E \parallel B$ and the negative LMC for $E \perp B$ provides the transport evidence for Weyl fermions and has been widely studied for both nonmagnetic and magnetic Weyl semimetal candidates[8, 21, 29, 79-82].

In this section, we discuss the Weyl-fermion-induced magnetotransport anomalies in Mn$_3$Sn and Mn$_3$Ge. The magnetoconductance obtained for Mn$_3$Sn at 60 K is strongly anisotropic (Fig. 5d)[8]. When the magnetic field $B$ is applied parallel to the electric field $E$, a positive magnetoconductance is observed, while a negative magnetoconductance appears in the $E \perp B$ configuration. In the low-field regime, the positive LMC for $E \parallel B$ shows a linear increase with $B$, consistent with the chiral anomaly of a type-II WSM[83, 84]. In magnetic conductors, the external magnetic field generally suppresses the charge carrier scattering due to thermal spin fluctuations, leading to a positive magnetoconductance irrespective of the angle between $E$ and $B$. While this effect may complicate the identification of the chiral anomaly, especially at high temperatures, such positive isotropic LMC should cease on cooling. In contrast, the positive LMC observed in Mn$_3$Sn for $E \parallel B$ rises dramatically with decreasing temperature from 300 K down to 60 K before entering into a cluster glass phase at $T < 50$ K (Fig. 5e)[8]. This behavior excludes the suppression of spin fluctuations or weak-localization as possible causes for the observed positive LMC, verifying that the chiral anomaly of the Weyl fermions is the primary driver of the significantly anisotropic magnetotransport in Mn$_3$Sn.

Unlike the case of Mn$_3$Sn, the ferro-octupole phase in Mn$_3$Ge remains intact further down to 0.3 K, which provides an advantage for discerning the intrinsic effects of Weyl fermions from that of the magnetic spin fluctuations. Figure 5f shows the field dependence of the magnetoconductivity $\Delta\sigma(B) = (\sigma(B) - \sigma(0))$ observed in Mn$_3$Ge at 0.3 K. The sharp dip in the magnetoconductivity for $B \leq 2$ T is most likely caused by magnetic domain effects, as the magnetization curve shows hysteretic behavior in the same field region at 2 K. The domain walls are most likely chiral and their effects on transport is an intriguing subject for future studies[11]. For $B > 2$ T, the magnetoconductivity measured in $B \parallel [01\bar{1}0]$ becomes negative in the $E \perp B$ configuration, indicating that the effect due to spin fluctuations is fully suppressed by lowering the temperature to approximately one-thousandth of the Néel temperature. In contrast, the magnetoconductivity in the $E \parallel B$ configuration is positive and is nearly quadratic in $B$ (Fig. 5f), which is distinct from the $B$-linear behavior observed in Mn$_3$Sn.

The $B^2$ dependence of the positive LMC is likely a result of strongly titled type-II Weyl cones with the applied magnetic field perpendicular to the tilt axis[84].

Apart from the chiral anomaly, other conventional mechanisms may yield positive LMC; one such well-known effect is the current jetting[85]. High-mobility semimetals may exhibit a very large transverse magnetoresistance (TMR) relative to the LMR owing to the orbital effect. In such cases, the strongly anisotropic MR may restrict the current to the direction parallel to the field, while the current flow transverse to the field is largely suppressed. This field-induced steering of the current may lead to positive LMC that depends strongly on the sample dimension and geometry. Recent studies have shown that the current jetting effect serves as the primary mechanism behind the positive LMC observed in several Dirac and Weyl semimetal candidates[85]. The carrier mobility of $Mn_3X$ ($X$ = Sn, Ge) is considerably lower than that of weakly correlated semimetals and zero-gap semiconductors, with a nearly isotropic magnetoresistance. Therefore, any current-jetting effect is unlikely to occur. Nevertheless, experimental tests were performed to address this concern. The magnetoconductivity measured at different voltage contacts on the sample are nearly identical, and all measured samples display qualitatively the same behavior, indicating that the observed positive LMC of $Mn_3X$ ($X$ = Sn, Ge) is intrinsic rather than arising from current inhomogeneities.

Another magnetotransport signature of the chiral anomaly is the planar Hall effect (PHE), which is formulated as

$$\Delta \sigma = \Delta \sigma_{\text{chiral}} \cos^2 \theta \qquad (5)$$

$$\Delta \sigma_H^{\text{PHE}} = \Delta \sigma_{\text{chiral}} \sin \theta \cos \theta \qquad (6)$$

where $\Delta\sigma$ and $\Delta\sigma_\text{H}^\text{PHE}$ are the longitudinal magnetoconductivity and the planar Hall conductivity, respectively[86]. Unlike the conventional Hall effect measurements with applied *E*- and *B*-fields being mutually perpendicular, the *E*- and *B*-fields in the PHE measurement are coplanar during the field angle ($\theta$) rotation, with the *B*-field lying within the kagome plane (see schematic in Fig. 5g). According to Eq. (5), the PHE reaches extrema at 45 ° and 135 °, and its amplitude is given by the chiral-anomaly induced positive magnetoconductivity. The angular dependence of $\Delta\sigma$ and $\Delta\sigma_\text{H}^\text{PHE}$ observed in Mn$_3$Sn are well described by the theoretical forms for the chiral anomaly (Fig. 5g), and further verifies the presence of magnetic Weyl fermions.

**Scaling Behavior in Mn$_3$*X* (*X* = Sn, Ge)**

The results summarized above provide firm evidence for the existence of Weyl nodes near the Fermi energy, which play a central role in generating the AHE and ANE. When the AHE comes from the intrinsic mechanism, the Hall conductivity is independent of the longitudinal conductivity, following the universal scaling law[26, 87-89]. Figure 6a presents the transverse Hall conductivity, $\sigma_\text{H}$, as a function of the longitudinal conductivity, $\sigma$, for various single crystals of Mn$_{3+x}$*X*$_{1-x}$ (*X* = Sn, Ge) with different composition in comparison with those of ferromagnetic materials. Here, $\sigma$ and $\sigma_\text{H}$ refer to the values obtained at the low temperature limit of the phase where the inelastic scattering is minimized. Notably, for all studied samples, the longitudinal $\sigma$ is in the range of $5 \times 10^3 \sim 10^4$ $\Omega^{-1}\text{cm}^{-1}$, and the $\sigma_\text{H}$ value is nearly independent of $\sigma$ (Fig. 6a inset). Both features indicate that Mn$_3$*X* (*X* = Sn, Ge) sits in the "good metal" regime, where the AHE is governed by the intrinsic mechanism.

**Summary and Perspectives**

The combination of the experimental and theoretical studies on the chiral antiferromagnets Mn$_3$X (X = Sn, Ge) has revealed a comprehensive set of evidence that confirms the presence of Weyl fermions, in particular, the large Berry-curvature-driven AHE and ANE, positive LMC and PHE originated from the chiral anomaly. The large Berry curvature stemming from the Weyl nodes leads to significantly enhanced AHE and ANE in the absence of net magnetization. To further demonstrate this feature, we show a summary comparison of the magnetization dependence of the spontaneous Hall conductivity $\sigma_\text{H}(M)$ and the anomalous Nernst coefficient $S_{yx}(M)$ obtained in Mn$_3$X (X = Sn and Ge) with those observed for various conventional ferromagnetic materials (Figs. 6b and c)[4-6, 9, 15, 21, 23, 89]. The large $\sigma_\text{H}(M)$ and $S_{yx}(M)$ of Mn$_3$X manifest themselves beyond the linear relation for the ordinary ferromagnets (red shaded region). Other known members of this regime are the ferromagnetic cobalt-based Weyl semimetals Co$_2$MnGa[21] and Co$_3$Sn$_2$S$_2$[23]; in both compounds, recent ARPES studies reveal evidence of Weyl fermions[74, 75], which are identified as the dominant cause of the unusually large AHE and ANE.

The research on Mn$_3$X (X = Sn, Ge) therefore paves the path for designing and investigating new Weyl magnets with strong electron correlation. The realization of Weyl magnets triggers further exploration of their intriguing properties emerging from the interplay between magnetism and topology, which may cast new light on the fundamental science behind topological quantum states of matter. The great tunability and robustness of Weyl fermions, combined with the unique advantages of the antiferromagnetic spin structure over ferromagnetic counterparts, render Mn$_3$X (X = Sn, Ge) remarkable potential to spark advances in spintronics and energy harvesting technologies. Intensive efforts are underway to make the novel properties of Weyl magnets useful for future applications[16-18, 20-22].

**Methods**

**Single crystal growth.** Single crystals of $Mn_3Ge$ were synthesized by the Bridgman and Bi self-flux methods, using high-purity starting materials (Bismuth-99.999%, Manganese and Germanium-99.999%) with an optimized ratio. The Bridgman growth follows a procedure similar to that described in previous studies of $Mn_3Sn$[9]. The initial step was to obtain polycrystalline samples by melting mixtures of manganese and germanium in an arc furnace under argon atmosphere. Starting from these polycrystalline materials, the single crystals of $Mn_3Ge$ were grown using a single-zone Bridgman furnace, by applying a maximum temperature of 1080 ℃ and a growth speed of 1.5 mm/h. In the case of Si-doped samples, the mixed germanium and silicon are melted with pure manganese in an arc furnace with the stoichiometric ratio. The subsequent Bridgman growth of single crystals was achieved with a lower maximum temperature of 1050 ℃ and a slower growth speed of 0.5 mm/h. The crystals were then annealed at 740 ℃ for 3 days.

**Crystal characterization.** The crystals were characterized by single-crystal X-ray diffraction (RAPID, Rigaku) at room temperature. The lattice parameters are obtained by Rietveld refinement. All the samples were shown to be single phase, with lattice parameters consistent with previous work[5]. According to the energy dispersive X-ray (EDX) analysis with a scanning electron microscope (SEM), the compositions of crystals obtained by Bi-flux and Bridgman methods are $Mn_{3.03}Ge_{0.97}$ and $Mn_{3.04}Ge_{0.96}$, respectively. The oriented single crystals were cut into a bar-shape by spark erosion for transport and magnetization measurements.

**Magnetization measurements.** The magnetization measurements on orientated samples were conducted using a commercial SQUID magnetometer (MPMS, Quantum Design) in the temperature range of 2 - 400 K.

**Electrical resistivity and Hall effect studies.** The longitudinal and Hall voltage signals were measured simultaneously in six-probe geometry; contacts to the crystals were made by gluing 20 μm gold wires with silver epoxy or attaching them by spot-welding. Measurements at temperatures of 2-400 K were performed in the PPMS system (Quantum Design), and a helium-3 sample-in-vacuum insert system (HelioxVT, Oxford Instruments) was employed for low-temperature measurements at 0.3 K. The Hall contributions to the longitudinal resistivity and vice versa were eliminated by adding and subtracting the resistivity data taken at positive and negative magnetic fields. The uncertainties of the longitudinal resistivity and Hall resistivity are about 1-2%.

**Thermoelectric studies.** The thermoelectric properties were measured by the one-heater and two-thermometer configuration using the PPMS system (Quantum Design). Bar-shaped samples with a typical dimension of 10 × 2 × 2 mm$^3$ were used for the measurements. By applying a temperature gradient -$\nabla T$ parallel to the long direction of the sample, the thermoelectric longitudinal and transverse emf voltages $V_i$ and $V_j$ were obtained in an open circuit condition. The thermal gradient -$\nabla T$ was monitored by two thermometers, which were approximately ∼5 mm apart and were linked to the sample via strips of ∼0.5 mm-wide copper-gold plates. The magnitude of the transverse voltage $\nabla V$ was linearly related to the applied temperature difference $\nabla T$, which was typically set to be 1.5 and 2.0% of the sample temperature for Seebeck and Nernst measurements, respectively. The Seebeck coefficient $S_{ii}$ and Nernst signal $S_{ji}$ were derived as $S_{ii} = E_i / \nabla T$ and $S_{ji} = E_j / \nabla T$ where $E_i$ and $E_j$ are the longitudinal and transverse electric fields. The magnetic field dependence of the Nernst signal was obtained after removing the longitudinal component, which is roughly field-independent. The uncertainties of the Seebeck and Nernst signals are dominated by the uncertainties of the corresponding geometrical factors and are estimated to be 10-20%.

**Density functional theory (DFT) calculations of the band structure.** The electronic structure calculations were performed using the Quantum ESPRESSO package[90]. The exchange-correlation energy functionals were considered within the generalized-gradient approximation (GGA), following the Perdew-Burke-Ernzerhof scheme[91]. A $7 \times 7 \times 7$ $\boldsymbol{k}$-point grid and projector augmented wave (PAW) pseudopotentials[92] were applied for the calculations. The cut off energies of 80 Ry and 320 Ry were chosen for the wave functions and the charge density, respectively. By using the Wannier90 code[93-95], the Wannier basis set was constructed from the Bloch states obtained in the DFT calculation, which includes 292 bands. The Wannier basis mentioned above consists of localized (*s,p,d*) - character orbitals at each Mn site, and (*s,p*) -character orbitals at Ge site, which gives 124 orbitals/u.c. in total. The Berry curvature, along with the transverse thermoelectric conductivity[63], was calculated using a Wannier-interpolated band structure[94] with $70 \times 70 \times 70$ $\boldsymbol{k}$-point grid and additional adaptive $\boldsymbol{k}$-point grid of $3 \times 3 \times 3$ in regions where the Berry curvature is large.

**Thermoelectric conductivity.** To calculate the thermoelectric conductivity, we begin with the calculation of the anomalous Hall conductivity, $\sigma_{\mathrm{AH}}$, at finite temperature:

$$\sigma_{\mathrm{AH}} = \sigma_{\mathrm{AH}}^{I} + \sigma_{\mathrm{AH}}^{II}$$

$$\sigma_{\mathrm{AH}}^{I} = \frac{-\hbar}{2\pi V} \Sigma_{\mathrm{k}} \int_{-\infty}^{\infty} d\epsilon \frac{\partial f(\epsilon)}{\partial \epsilon} \mathrm{Tr}\big[\hat{\jmath}_X G^R \hat{\jmath}_Y G^A\big]$$

$$\sigma_{AH}^{II} = \frac{-\hbar}{4\pi V} \sum_k \int_{-\infty}^{\infty} d\epsilon f(\epsilon) \ \text{Tr}\left[\hat{J}_X \frac{\partial \hat{G}^R}{\partial \epsilon} \hat{J}_Y \hat{G}^R - \hat{J}_X G^R \hat{J}_Y \frac{\partial \hat{G}^R}{\partial \epsilon} - <R \to A>\right]$$

Here, $\hat{J}_X$ and $\hat{J}_Y$ denote the current operator in the *x* and *y* direction, and $\hat{G}^R$ and $\hat{G}^A$ represent the retarded and advanced Green's function, respectively. $V$ is the volume of the system. In the regarded Green's function, we consider the self-energy

$$\Sigma = \left(1 - \frac{1}{Z}\right)\epsilon - iA\epsilon^2 - i\gamma$$

We obtained the thermoelectric conductivity, $\alpha_{ji}$, following the Mott relation, that is

$$\alpha_{ji} = -\frac{\pi^2}{3|e|T}(k_B T)^2 \frac{\partial \sigma_{ji}(\epsilon)}{\partial \epsilon}$$

**Acknowledgements**

This work is partially supported by CREST (JPMJCR18T3), Japan Science and Technology Agency, by Grants-in-Aids for Scientific Research on Innovative Areas (15H05882, 15H05883, and 15K21732) from the Ministry of Education, Culture, Sports, Science, and Technology of Japan, by New Energy and Industrial Technology Development Organization, and by Grants-in-Aid for Scientific Research (19H00650) from the Japanese Society for the Promotion of Science (JSPS). The work at the Institute for Quantum Matter, an Energy Frontier Research Center was funded by DOE, Office of Science, Basic Energy Sciences under Award # DE-SC0019331. The work for first-principles calculation was supported in part by JSPS Grant-in-Aid for Scientific Research on Innovative Areas (18H04481 and 19H05825) and by MEXT as a social and scientific priority issue (Creation of new functional devices and high-performance materials to support next-generation industries) to be tackled by using post-K computer (hp180206 and hp190169). The use of the facilities of the Materials Design and Characterization Laboratory at the Institute for Solid State Physics, The University of Tokyo, is gratefully acknowledged.

**Author contributions**

S.N. conceived the project and planned the experiments. T.C., T.T., M.F., and S.N. performed the experiments and analyzed the data. S.M., T.K., F.I., and R.A. performed the first-principles calculations. D.N and R.I. performed the TEM and ICP measurements and analyses. S.N., M.F., T.T., S.M. and T. C wrote the paper; All authors discussed the results and commented on the manuscript.

**Correspondence**

Correspondence and requests for materials should be addressed to S.N. (email: satoru@phys.s.u-tokyo.ac.jp).

**Competing interests**

The authors declare no competing interests.

**Data availability statement**

The data that support the plots within this paper and other finding of this study are available from the corresponding author upon reasonable request.

**Figure 1 Magnetic structure and thereotically prediced Weyl points of $Mn_3X$ ($X$ = Sn, Ge) a,** Crystal structure of $Mn_3X$ ($X$= Sn, Ge) and the spin texture under a magnetic field applied along the [2$\bar{1}\bar{1}$0] direction. The Mn moments lying within the kagome (hexagonal *ab*) plane form the inverse triangular spin order (colored in purple). Here, we define [2$\bar{1}\bar{1}$0], [01$\bar{1}$0], and [0001] as the *x*, *y*, and *z* axes, respectively. The colored atoms lie within the kagome plane at $z = 0$. The large (small) spheres in grayscale represent Mn (Sn or Ge) atoms forming kagome planes at $z = \pm 1/2$. **b,** Schematic illustration of the cluster magnetic octupoles formed under an applied magnetic field $B \parallel y$. The arrows represent Mn magnetic moments that constitue the inverse triangular spin structure. The Mn moments on an octahedron made of two triangular units of adjacent kagome layers can be characterized by a cluster octupole moment (colored hexagons). The spin structure can then be considered as a ferroic ordering of cluster magnetic octupoles. **c,** The calculated band structure along the K-M-K cut (solid lines) of Mn$_3$Sn. The open and closed arrows respectively mark the opposite chirality (+) and (-) of the two low-energy Weyl point pairs, W$_1$ and W$_2$ (see the panel **d**). The studied Mn$_3$Sn samples are off-stoichiometric Mn$_{3.03}$Sn$_{0.97}$ and Mn$_{3.06}$Sn$_{0.94}$; the extra Mn shifts the chemical potential up in energy by 19 meV for Mn$_{3.03}$Sn$_{0.97}$ (red dashed line) and by 40meV for Mn$_{3.06}$Sn$_{0.94}$ (blue dashed line) compared to the $E_F$ of the stoichiometric sample (black solid line). **d,** Location of Weyl points in the $k_z = 0$ plane of the hexagonal Brillouin zone (BZ) for electron bands near $E_F$. Two groups of Weyl points W$_1$ and W$_2$ are shown. For each pair, the open and closed circles indicate positive (+) and negative (-) chirality, respectively. Rotation of an applied magnetic field within the *xy* plane may shift the position of the Weyl nodes along the hypothetical nodal ring (dashed circular curves). The present distribution of Weyl points corresponds to the magnetic structure of Mn$_3$Sn shown in **a**. **e,** Calculated surface local density of state (LDOS) at $E_F$ of Mn$_3$Ge that reveals the pair of Weyl points W$_1$ of oppsoite chirality (white and black dots) and the Fermi arcs marked as p$_1$ and p$_2$. Adapted from ref.[8], Springer Nature (**c,d**); and ref.[28], IOP (**e**).

**Figure 2 Observation of large anomalous Hall and Nernst effects in Mn$_3$Sn. a,** Field dependence of the Hall resistivity, $\rho_H(B)$, observed in Mn$_{3.02}$Sn$_{0.98}$ at room temperature under magnetic fields $B \parallel$ [2$\bar{1}\bar{1}$0], [01$\bar{1}$0], and [0001]. **b,** Anisotropic behavior of the magnetization *M* of Mn$_{3.02}$Sn$_{0.98}$ observed at room temperature for the three different field directions. **c,** Temperature dependence of the zero-field Hall conductivity, $\sigma_H(B=0)$, in Mn$_{3.02}$Sn$_{0.98}$. **d,** Anisotropic field dependence of the Nernst coefficient $-S_{ji}$ of Mn$_{3.06}$Sn$_{0.94}$ (left-axis). The field dependence of the magnetization *M* for $B \parallel$ [2$\bar{1}\bar{1}$0] (right-axis) is shown for comparison. **e,** Temperature dependence of the magnetization *M* of Mn$_{3.04}$Ge$_{0.96}$ obtained at $B = 0.1$ T along [2$\bar{1}\bar{1}$0], [01$\bar{1}$0], and [0001] directions, respectively. The magnetization is strongly anisotropic, such that the ratio between the in-plane and out-of-plane *M* reaches about 10 in the low temperature region. This feature is expected for the coplanar magnetic structure. Inset: The variation of the Néel temperature $T_N$ as a function of the Mn concentration *x*. The red and blue symbols represent $T_N$ values obtained in Mn$_{3+x}$Ge$_{1-x}$ and Mn$_{3+x}$(GeSi)$_{1-x}$, respectively. **f,** Temperature dependence of the Hall conductivity, $-\sigma_{zx}$, observed in Mn$_{3.03}$Ge$_{0.97}$ and Mn$_{3.04}$Ge$_{0.96}$. **g,** Anomalous Nernst coefficient, $-S_{yz}(B)$, (left-axis) vs. *B* obtained for $B \parallel$ [2$\bar{1}\bar{1}$0] and $Q \parallel$ [0001], at 100 K and 300 K. The isothermal magnetization *M* curves (right-axis) are also shown at the same temperatures, which exhibit approximately the same coercivity field of about 80 Oe as in $-S_{yz}(B)$. **h,** Temperature dependence of the zero-field Nernst coefficient, $-S_{zx}(T)$, for Mn$_{3.03}$Ge$_{0.97}$ and Mn$_{3.04}$Ge$_{0.96}$ obtained in field sweep measurements. Inset: Seebeck coefficients $S_{yy}(T)$ ($Q \parallel [01\bar{1}0]$) and $S_{zz}(T)$ ($Q \parallel [0001]$) of Mn$_{3.04}$Ge$_{0.96}$ as a function of $T$ measured under zero field. Both $S_{yy}(T)$ and $S_{zz}(T)$ retain negative values at low temperatures, and $S_{yy}(T)$ exhibits a pronounced negative peak at about 60 K. Adapted from ref.[4], Springer Nature (**a, b, c**) and ref.[9], Springer Nature (**d**).

**Figure 3 Anomalous transverse Wiedemann-Franz (WF) law found in Mn$_3$X ($X$ = Sn, Ge). a, b,** Field dependence of the thermal Hall conductivity $\kappa_{xz}$ (**a**) and $\kappa_{zy}$ (**b**). **c,** Temperature dependence of the anomalous Lorentz number $L_{zx}^{AH} = (e/k_B)^2 T \kappa_{zx}^A / \sigma_{zx}^A$ for Mn$_{3.06}$Sn$_{0.94}$ (filled blue squares) and Mn$_{3.09}$Sn$_{0.91}$ (filled red squared). The horizontal axis indicates the temperature normalized by the Debye temperature, $\theta_D$. The $\theta_D$ of the two samples is about 300 K. The dashed line marks $L_0 = \frac{\pi^2}{3}$. The $L_{xy}^{AH}$ of conventional ferromagnetic metals Ni with $\theta_D \approx$ 450 K (open purple circles) and Fe with $\theta_D \approx$ 470 K (open green triangles) are also included for comparison[67, 68]. Inset: Temperature dependence of $L_{xy}^{AH}$ observed in Ni and NiCu alloys[68]. The definition of $L_{zx}^{AH}$ and $L_0$ here is different from that in Eq. 3. **d,** Temperature dependence of the anomalous Hall conductivity, $-\sigma_{zx}^A$ (top), the anomalous thermal Hall conductivity divided by temperature, $-\kappa_{zx}^A/T$ (middle), and the anomalous Lorenz number, $L_{zx}^A = T\kappa_{zx}^A/\sigma_{zx}^A$ (bottom) for Mn$_3$Ge. Adapted from ref.[56], APS (**a, b**), ref.[66] (**c**), and ref.[13], AAAS (**d**).

**Figure 4 Transverse thermoelectric effects in Mn$_3$X ($X$ = Sn, Ge) and their theoretical analyses. a, b, c,** Temperature dependence of the Nernst coefficient, $-S_{ij}$, measured under $B$ = 14 T. $S_{zx}$ (red circle), $S_{yz}$ (blue diamond), and $S_{xy}$ (black triangle). The label definitions are modified from the original ones in ref.[15] so that they become consistent with the results obtained by other groups. (**a**). zero-field anomalous transverse thermoelectric conductivity, $-\alpha_{xz}$, with the best fit to the minimal model of Weyl fermions (solid line) (**b**), and the two contributions to $-\alpha_{xz}$ (**c**) in Mn$_3$Ge. **d,** Temperature dependence of $-\alpha_{xz}$ in Mn$_{3.06}$Sn$_{0.94}$ and Mn$_{3.09}$Sn$_{0.91}$ from ref.[9]. **e,** $k$-space distribution of the Weyl points (W$_1$-W$_9$) under $B \parallel [0\bar{1}10]$ predicted by the DFT calculation in ref.[30]. The red and blue frames mark the W$_1$ and W$_3$ pairs that are consistent with the WP1 (**f**) and WP3 (WP'3) (**g**) pairs revealed by our calculation. **f, g,** Contour plots of the theoretical Berry curvature spectrum $|\Omega_z|$ ($k_x$, $k_y$, $k_z$) in the $k_x - k_y$ plane under $B \parallel [01\bar{1}0]$ at $k_z$ = 0 (Å$^{-1}$) (**f**) and at $k_z$ = 0.137 (Å$^{-1}$) (**g**). The arrows indicate the Berry curvature arising from the WP1 (**f**) and WP3 (WP'3) (**g**) pairs. Each pair consists of two Weyl points with opposite chirality. **h, i,** Experimental $-\alpha_{zx}/T$ (solid symbols) versus theoretical curves obtained from the DFT calculations for Mn$_3$Sn (**h**) and Mn$_3$Ge (**i**). Calculations are made with the energy shifts in the chemical potential corresponding to the extra Mn in the off-stoichiometric samples, relative to the Fermi level $E_F$ of the stoichiometric case (i.e $E = E_F$). A convergence factor of $\gamma$ = 0.02 and a coefficient $AZ^2$ =5.0 eV$^{-1}$ are used in the calculation. The renormalized temperature $T^*$ = $T/5$ for Mn$_3$Sn is set accordingly to the ARPES results. A smaller renormalization factor $T^*$ = $T/2$ is chosen for Mn$_3$Ge given its larger bandwidth relative to that of Mn$_3$Sn. The inset of (**i**) shows

the calculated Hall conductivity $-\sigma_{zx}$ for $Mn_{3.06}Ge_{0.94}$ and $Mn_{3.03}Ge_{0.97}$ using the same parameters. Adapted from ref.[15], APS (**a, b, c, d**).

**Figure 5 ARPES study and the chiral anomaly in Mn$_3$X (X = Sn, Ge) a,** $k_z$ - dispersion along the H-K-H high-symmetry line (black arrows) obtained at 60 K by varying the incident photon energy $h\upsilon$ from 50 to 170 eV. The blue arrow marks the quasiparticle peak developed around K point. **b,** ARPES intensity at $E_F$ in the $k_x$–$k_y$ plane obtained with $h\upsilon$ = 103 eV. The solid curves represent the Fermi surface predicted by the DFT calculation. **c,** ARPES intensity maps in the $E$ - $k_x$ plane along the high-symmetry K-M-K direction at various $k_y$ obtained at 60 K. The original spectrum intensity (left) are compared to the intensity divided by the energy-resolution convoluted Fermi-Dirac (FD) function. The solid lines represent the calculated band dispersion, with the red and blue curves marking the electron and hole bands that form the Weyl points. Right: The anomalies in the corresponding momentum distribution curves (MDCs) are marked by the vertical color bars. **d,** Field dependence of the magnetoconductivity in Mn$_3$Sn for $B \parallel I$ (red) and $B \perp I$ (blue) under a magnetic field $B \parallel [01\bar{1}0]$ at 60 K. **e,** Field dependence of the magnetoconductivity in Mn$_3$Sn for $B \parallel I$ measured at various temperatures. **f,** Field dependence of the magnetoconductivity for $B \parallel I$ and $B \perp I$ and with $I \parallel [0001]$ measured at 0.3 K in Mn$_3$Ge. The dashed line is a fit to the $B^2$ dependence. **g,** Angle dependence of the longitudinal magnetoconductivity (top) and the planar Hall effect (PHE) (bottom) in $Mn_{3.06}Sn_{0.94}$ measured at 300 K and 100 K under $B$ = 3 T. The solid lines are the fits to the theological forms (see the main text for details). Adapted from ref.[8], Springer Nature (**a, b, c, d, e**).

**Figure 6 Universal scaling relations for the magnetic Weyl semimetals. a,** Universal scaling relation between the Hall conductivity, $\sigma_H$, and the longitudinal conductivity $\sigma$ for $Mn_{3.04}Ge_{0.96}$ (2 K and 10 K), $Mn_{3.06}Sn_{0.94}$ (60 K and 100 K), and $Mn_{3.09}Sn_{0.91}$ (100 K)[4, 9], plotted together with the data for various ferromagnets including transition metals (Ni, Gd, Fe, and Co thin films)[87], Fe single crystals[87], Co$_3$Sn$_2$S$_2$[31], MnSi[89], Fe$_{1-x}$Co$_x$Si[89], perovskite oxides (SrRuO$_3$, La$_{1-x}$Sr$_x$CoO$_3$, La$_{1-x}$(SrCa)$_x$MnO$_3$)[71], spinels (Cu$_{1-x}$Zn$_x$Cr$_2$Se$_4$)[71], pyrochlore (Nd$_2$(MoNb)$_2$O$_7$)[71], and magnetic semiconductor (Ga$_{1-x}$Mn$_x$As, In$_{1-x}$Mn$_x$As, anatase- Co-TiO$_2$, rutile- Co-TiO$_2$)[71]. Within the good metal regime of $3 \times 10^3 \leq \sigma \leq 5 \times 10^5$ $\Omega^{-1}cm^{-1}$ (yellow shaded area), $\sigma_H$ is predominated by the intrinsic Berry phase contribution. The value of $\sigma_H$ lies below about $10^3 \Omega^{-1}cm^{-1}$ (horizontal dotted line) and is nearly constant with $\sigma$. In the regime of $\sigma \geq 5 \times 10^5$ $\Omega^{-1}cm^{-1}$, the extrinsic skew scattering contribution dominates $\sigma_H$. The solid line presents the theoretical prediction by ref[88]. Inset: Zoomed plot showing the data for the Mn$_3$Sn and Mn$_3$Ge single crystals with different compositions, which corresponds to the region framed by the red rectangle in the main panel. **b, c,** Full logarithmic plot of the magnetization ($M$) dependence of the Nernst coefficient $|S_{ji}|$ (**b**) and the Hall conductivity $\sigma_H$ (**c**) for Mn$_3$Sn (blue), Mn$_3$Ge (red), and other Weyl magnets[,4, 9, 21, 31], in comparison with those for ordinary ferromagnets[9, 26, 71, 87, 96]. The red shades mark the regions in which $|S_{ji}|$ or $\sigma_H$ is linearly related to $M$ for conventional ferromagnets. The Weyl magnets that violate this linear relation are located in the blue shaded region.

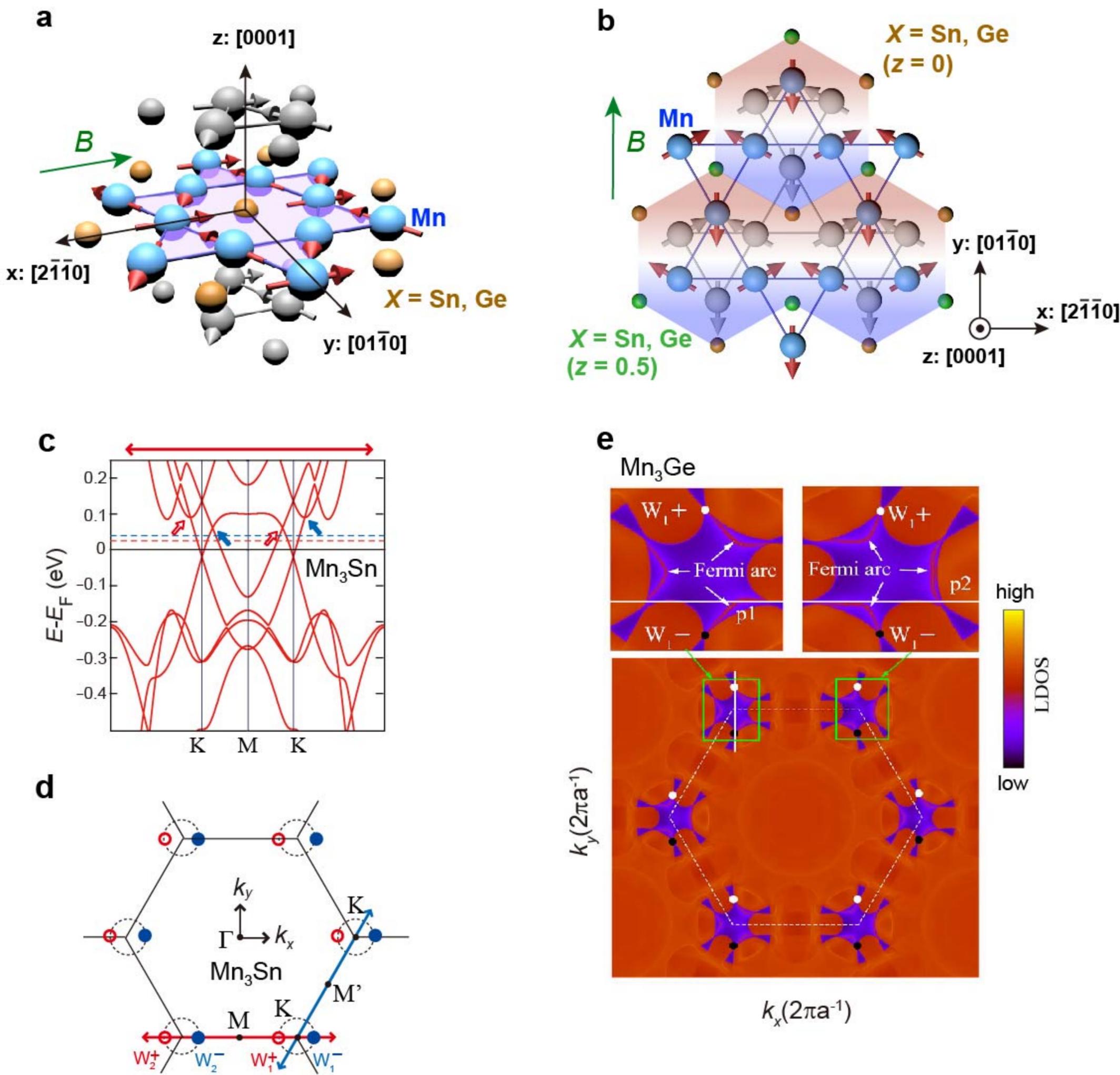

Figure 1

# Mn₃Sn

# Mn₃Ge

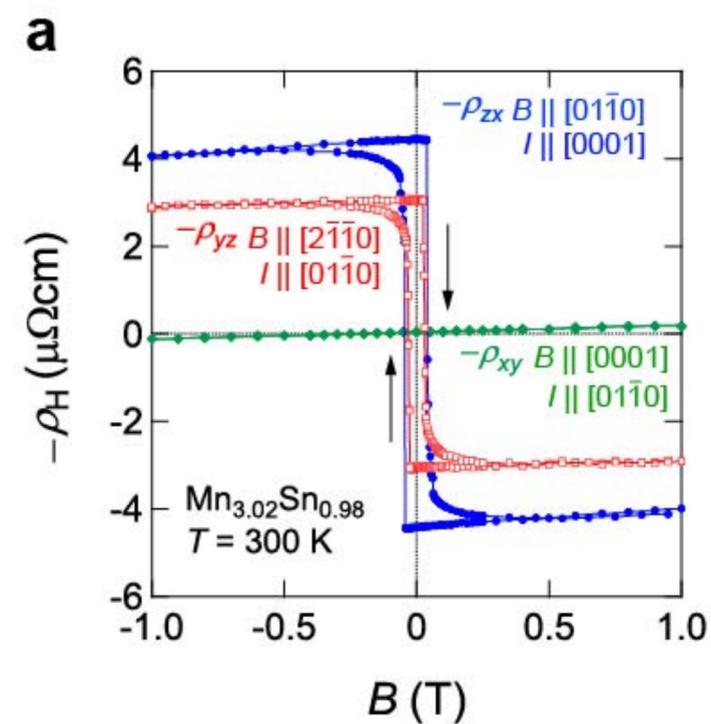
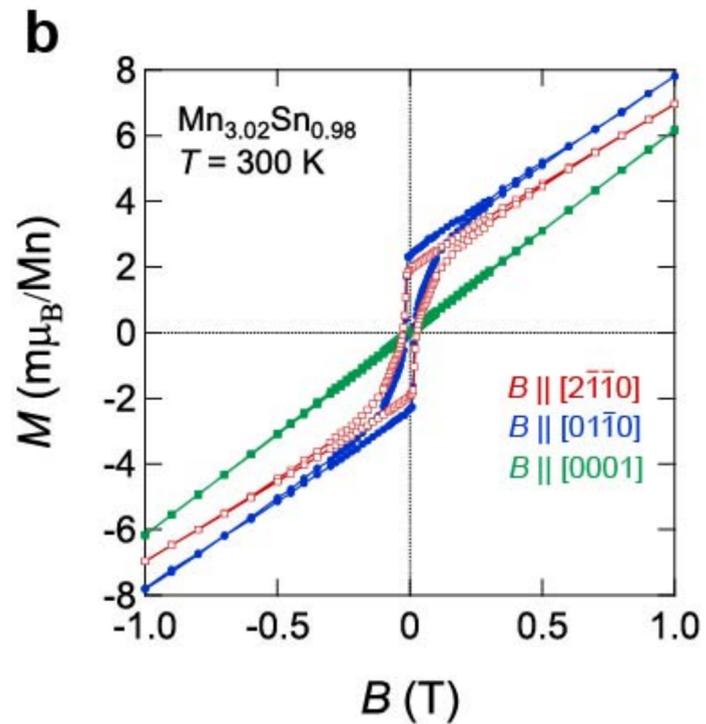
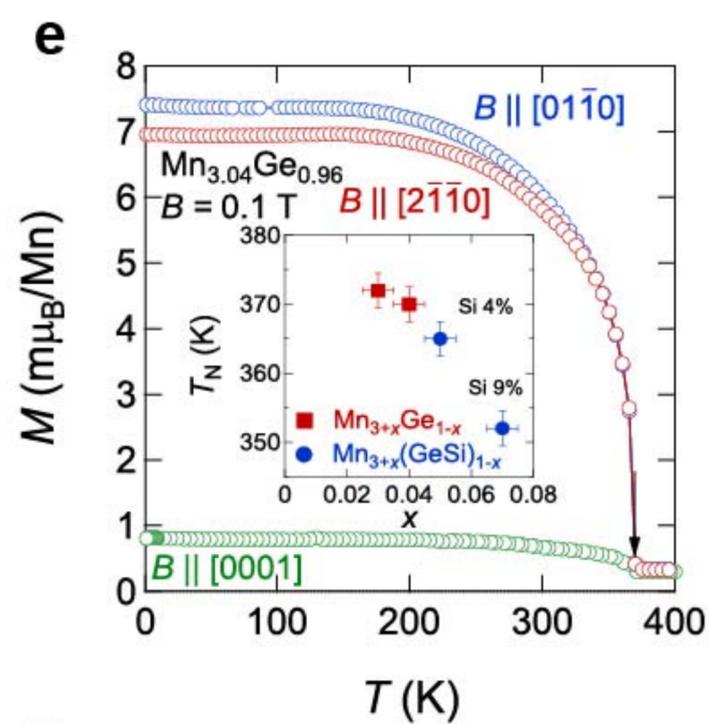
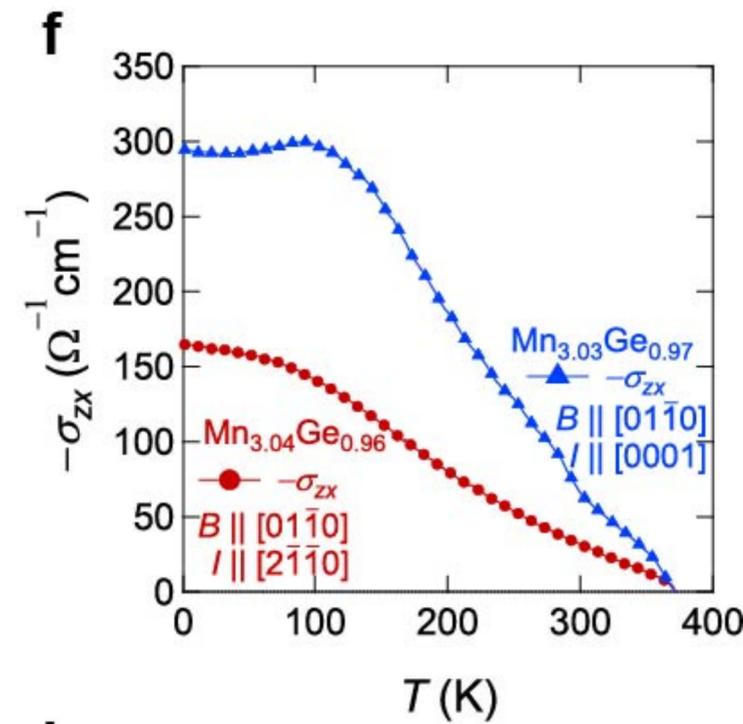
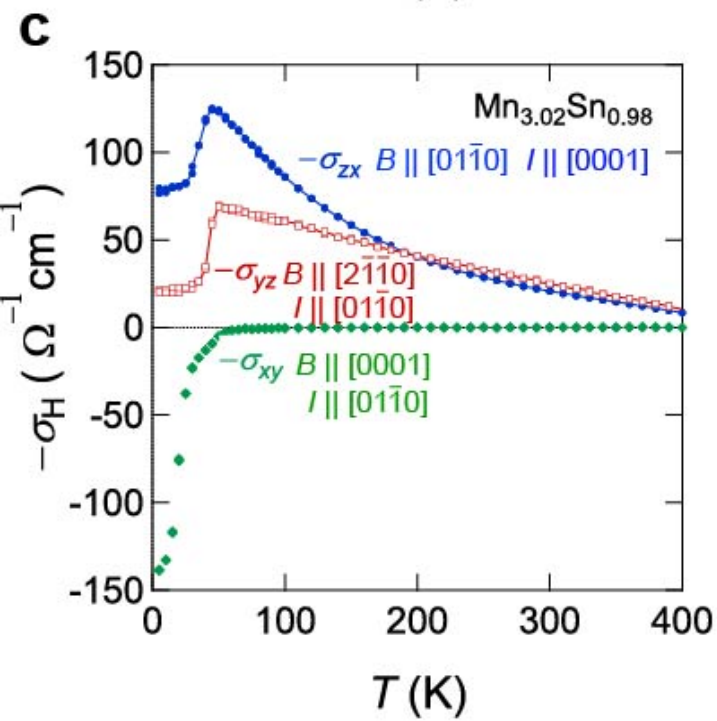
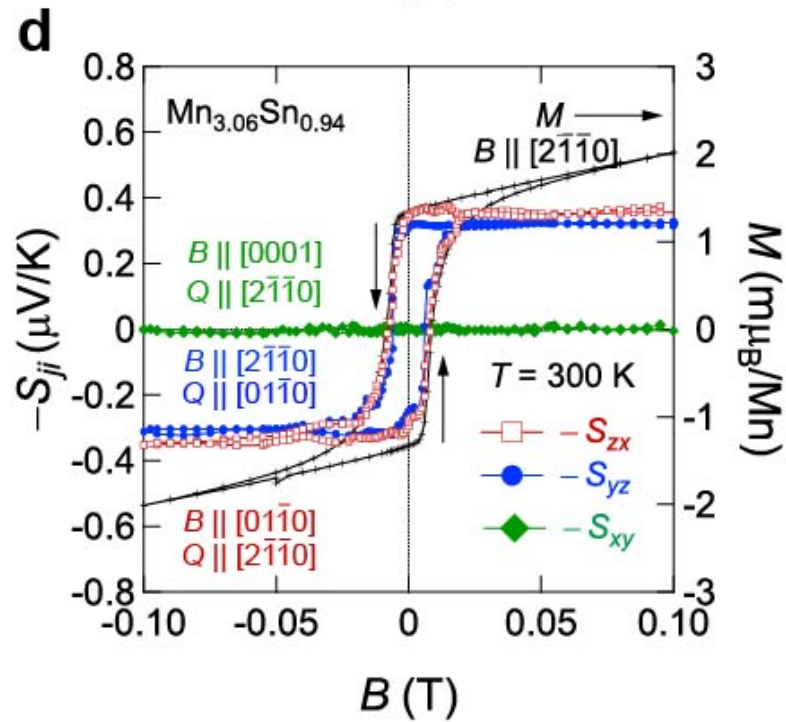
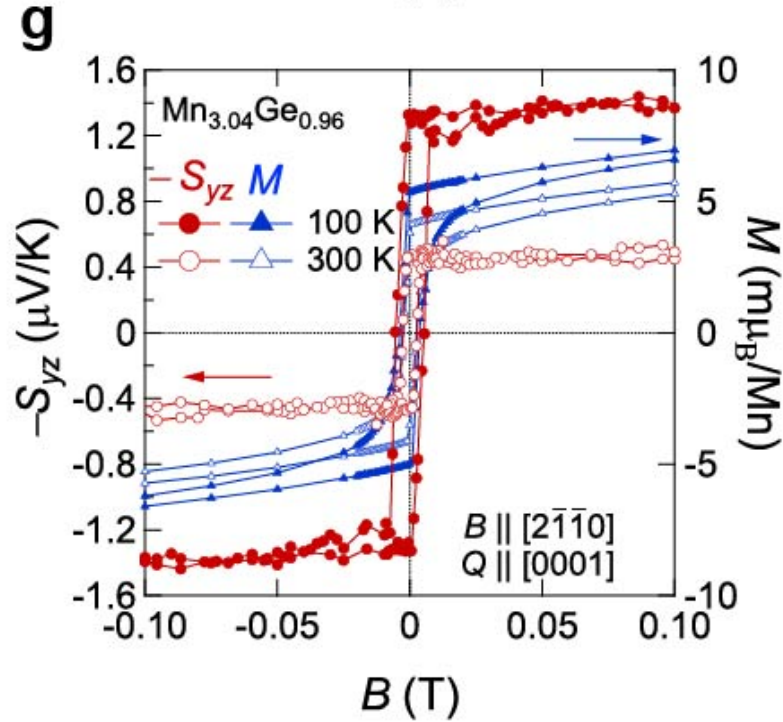
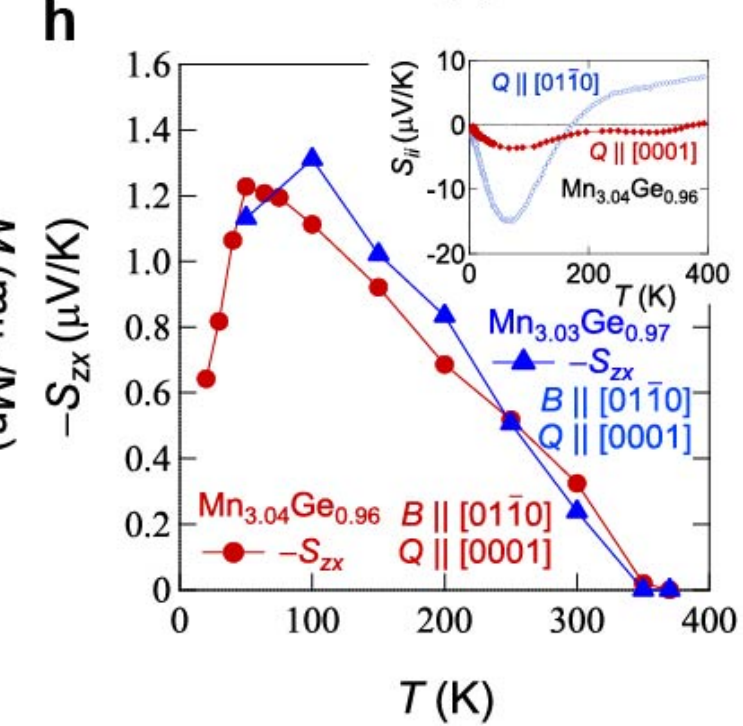

Figure 2

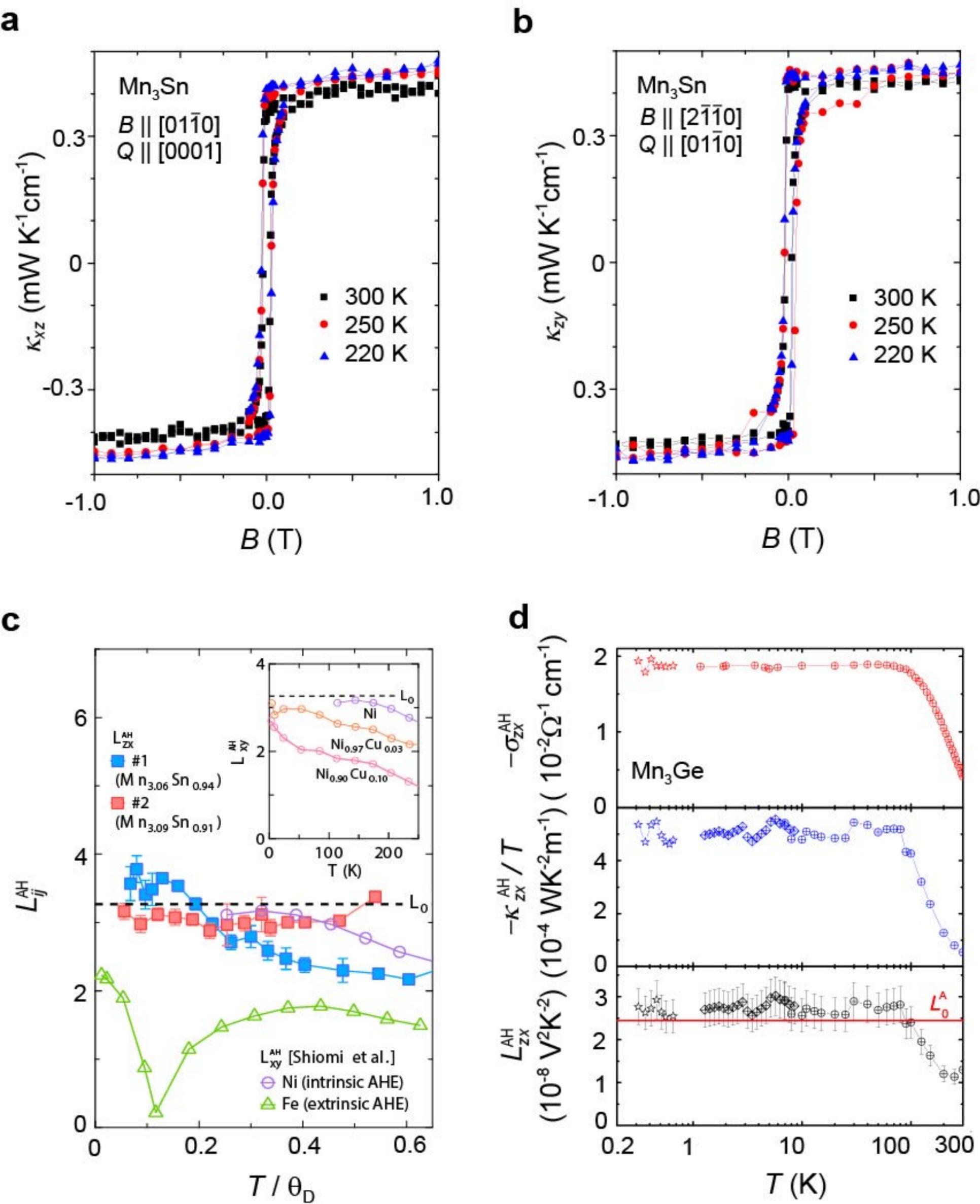

Figure 3

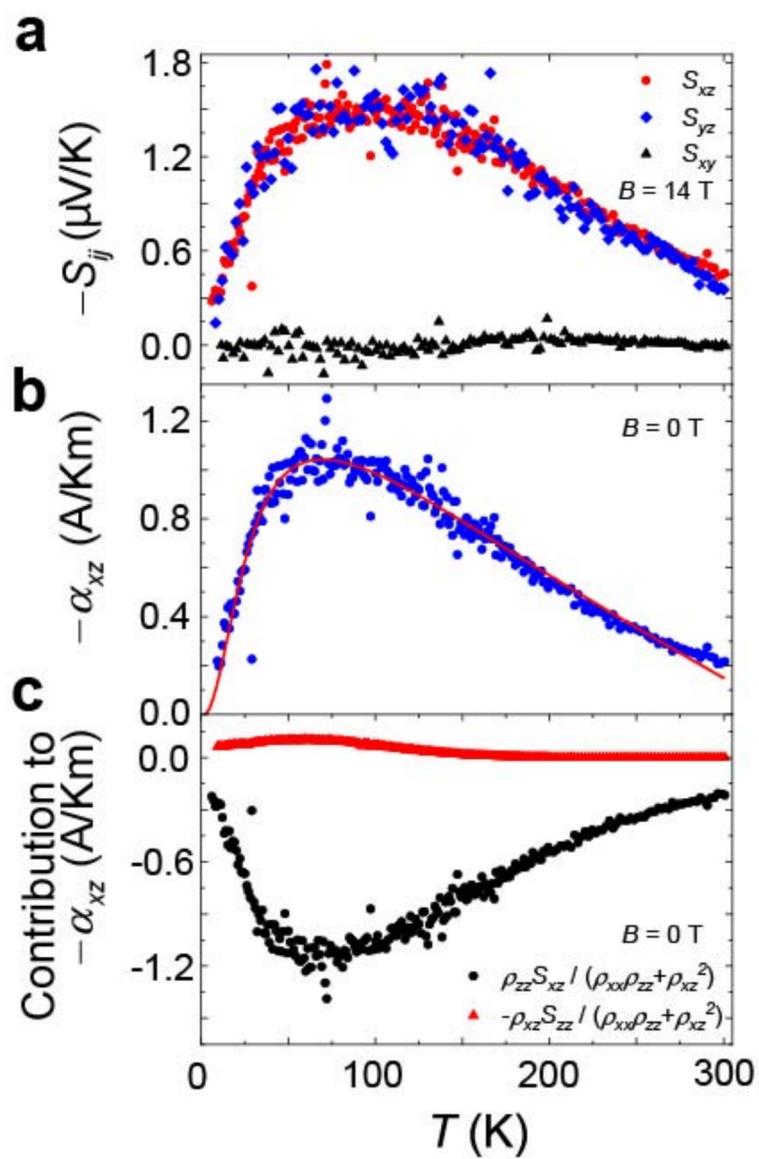
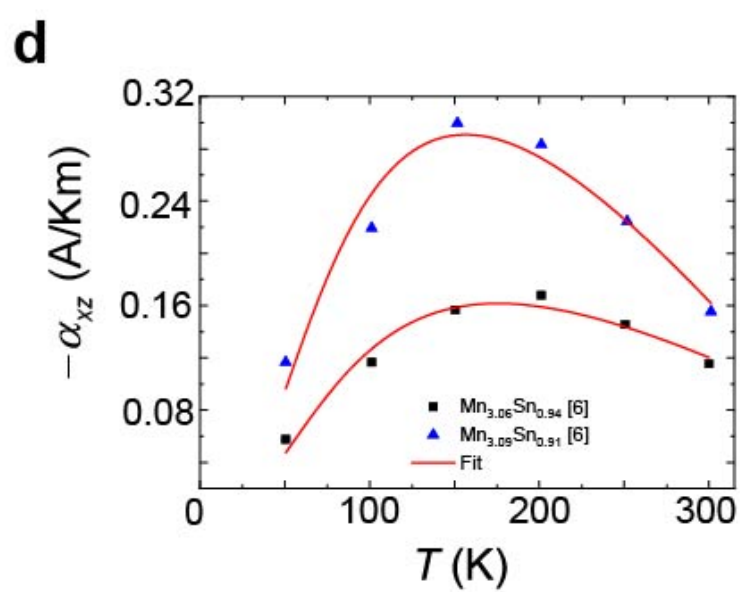
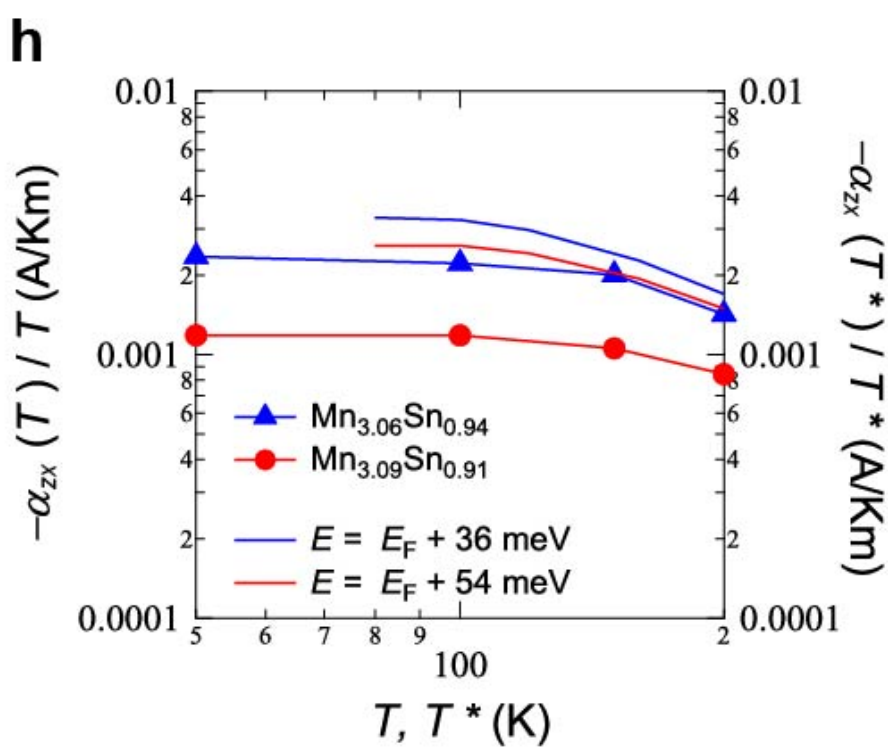
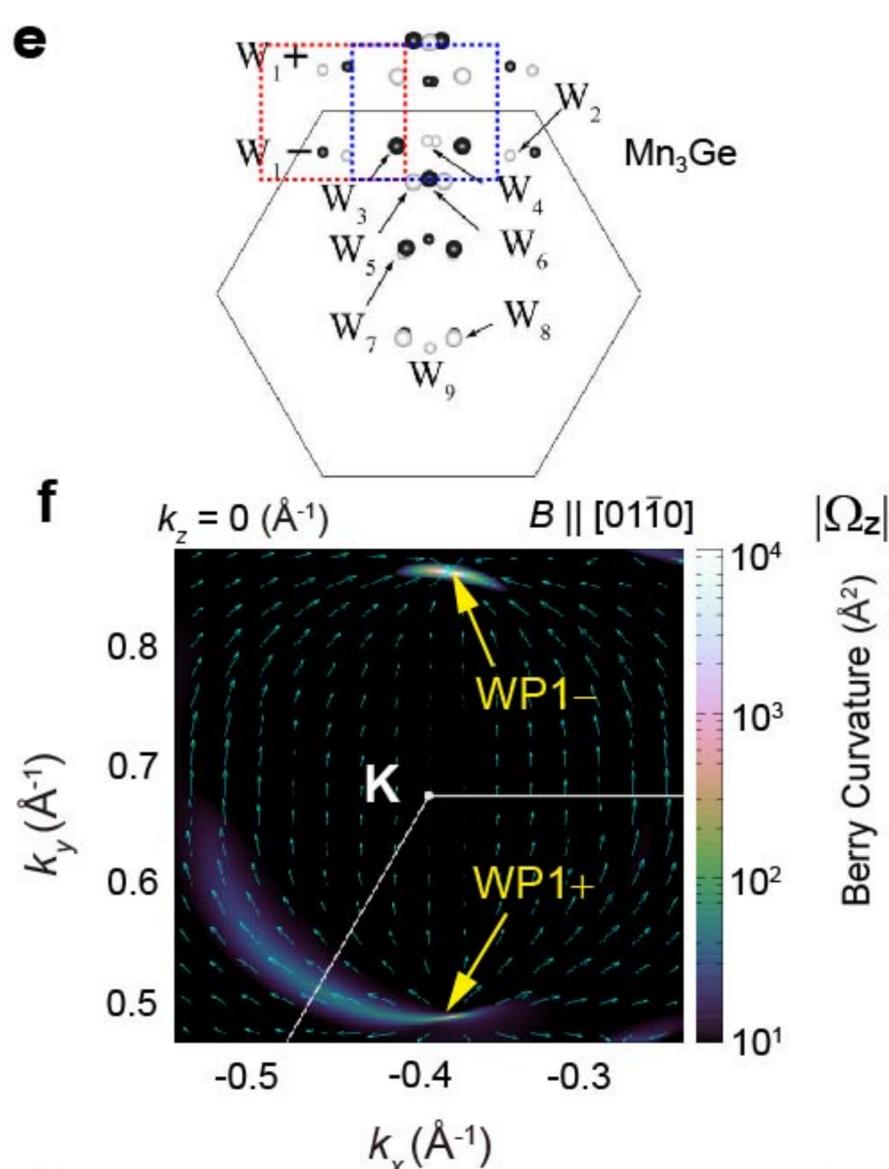
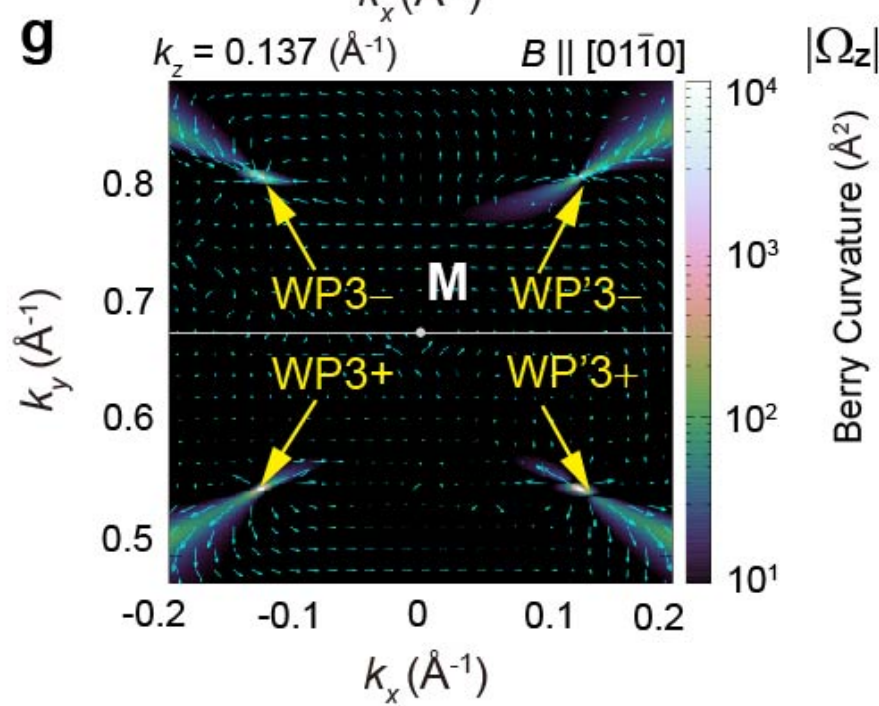
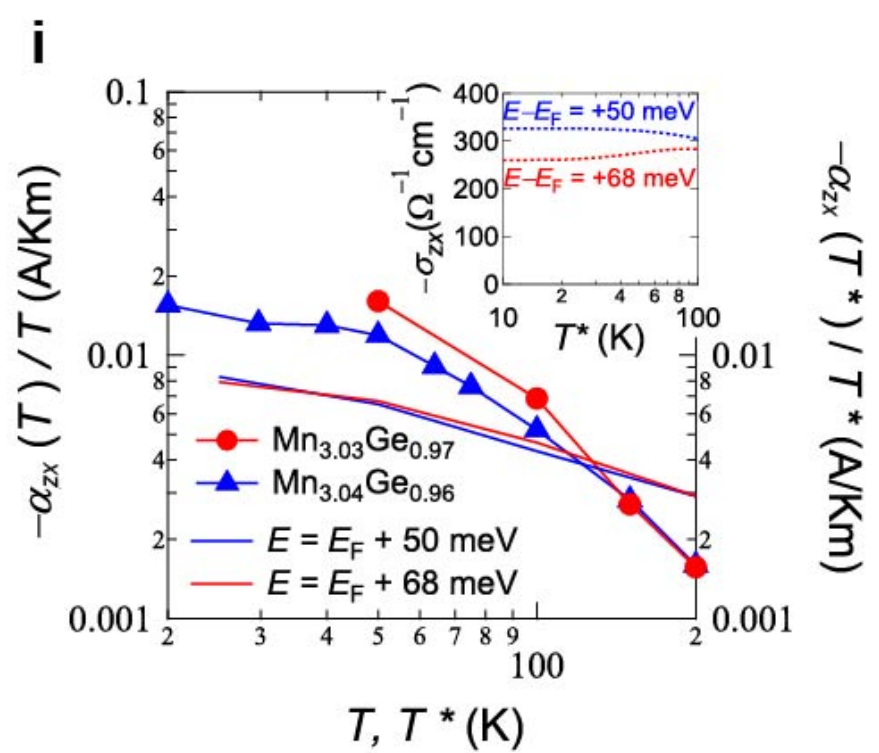

Figure 4

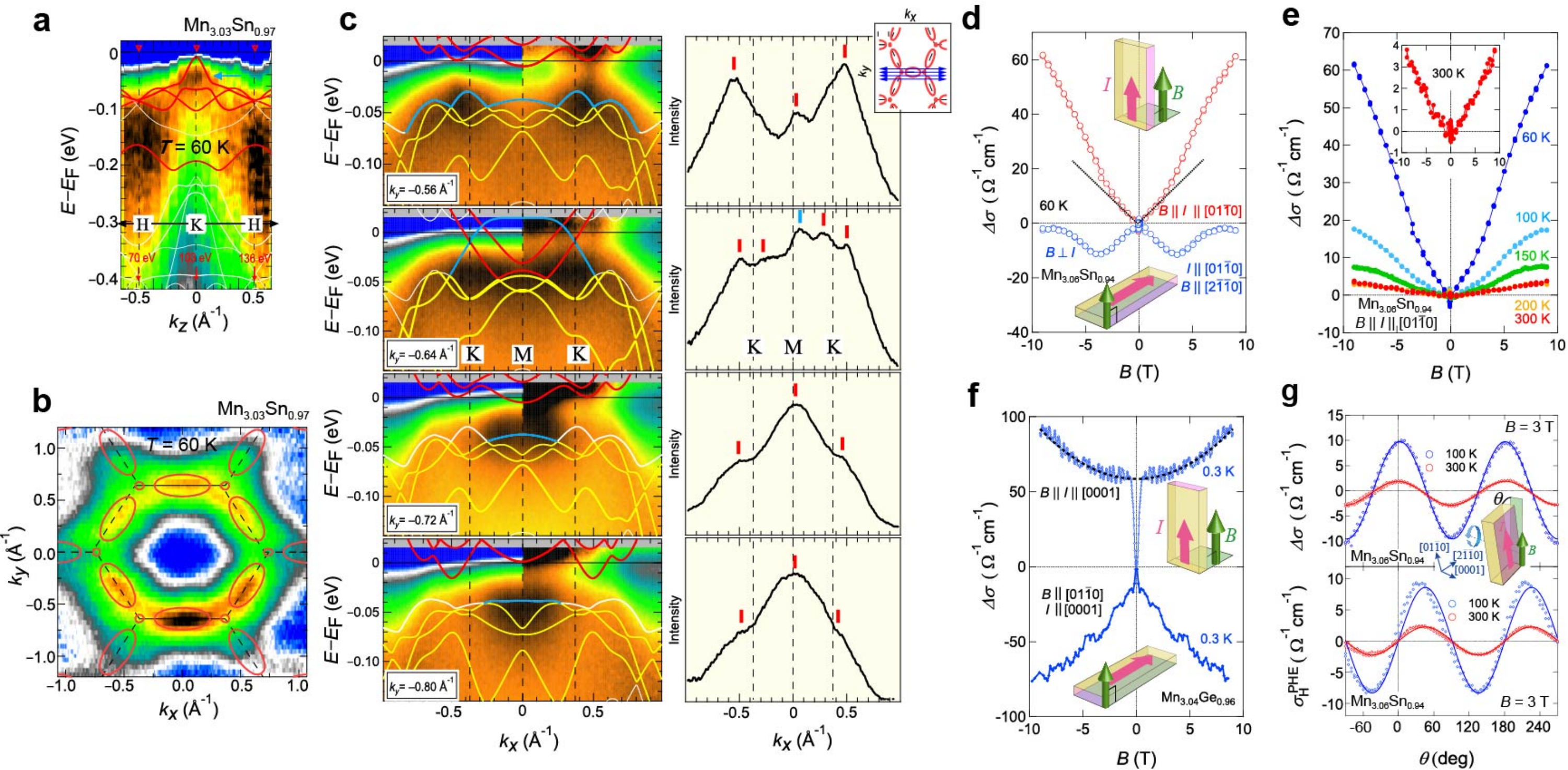

Figure 5

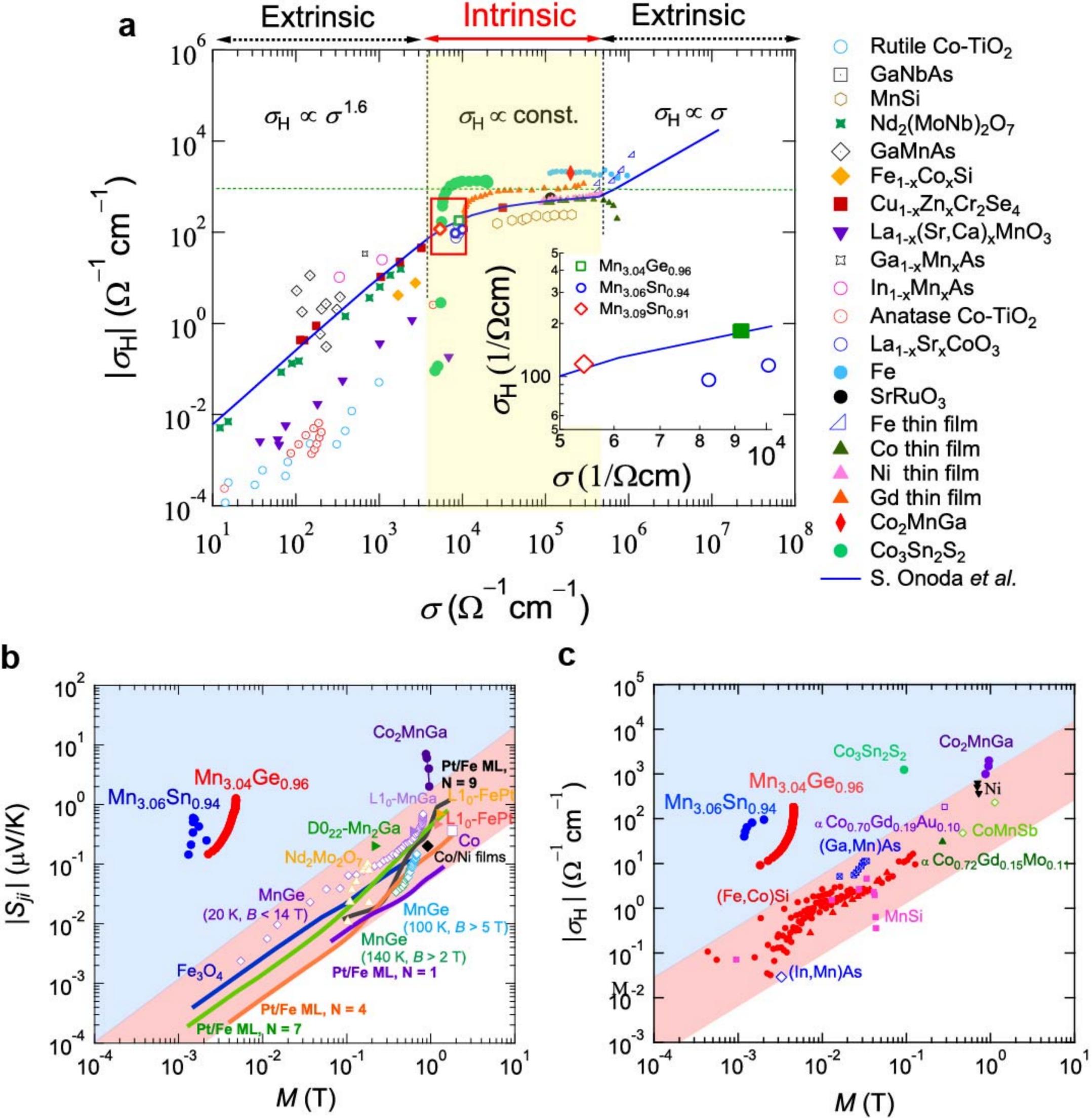

Figure 6